\documentstyle[aps,prl]{revtex}
\title{Binary Mixtures of Bose-Einstein Condensates: 
Phase Dynamics and Spatial Dynamics}
\author{Alice Sinatra and Yvan Castin \\
Laboratoire Kastler Brossel
\thanks{Unit\'e de recherche de  l'Ecole normale sup\'erieure et de
l'Universit\'e Pierre et Marie Curie, associ\'ee au CNRS.} \\
24 rue Lhomond, 75231 Paris Cedex 5, France}

\date{April 22, 1999}
\begin{document}
\maketitle

\begin{abstract}
We investigate the relative phase coherence properties  and
the occurrence of demixing instabilities
for two mutually interacting and time evolving Bose-Einstein condensates
in traps. Our treatment naturally includes the additional
decoherence effect due to fluctuations in the total number of particles.
Analytical results are presented for
the breathe-together solution, an extension of previously known
scaling solution to the case of a binary mixture of condensates.
When the three coupling constants 
describing the elastic interactions among the atoms in
the two states are close to each other, a dramatic 
increase of the phase coherence time is predicted.
Numerical results are presented for the parameters
of the recent JILA experiments.
\end{abstract}

\section{Introduction}
Since the recent experimental observation of Bose-Einstein condensation
in dilute atomic gases \cite{BEC}, 
much interest has been raised about the coherence
properties of the condensates. Considerable attention has been devoted to
the matter of the relative phase between two condensates: how this
phase manifests  itself in an interference experiment 
\cite{Javanainen,Ketterle}, how it
can be established by measurement \cite{Zoller,Jean}, 
and how it evolves in presence of atomic
interactions \cite{Jean,Walls,Maciek} and in presence of particle losses 
\cite{resur}.

As it was proved in recent experiments performed at JILA,
binary mixtures of condensates
represent an ideal system to study the phase coherence properties of
Bose-Einstein condensates \cite{JILA_phase}.
In these experiments two condensates in two different internal atomic
states are created with a well defined relative phase.
After a time $\tau$ during which the condensates evolve
in the trapping potentials, one mixes coherently
the two internal atomic states
which makes the two condensates interfere; from the spatial interference
pattern one gets
the relative phase of the two condensates. By repeating the whole experimental
process, one has access to the distribution of the relative phase
after an evolution time $\tau$, so that
one can investigate phase decoherence as function of time.

The interaction between the two
condensates in the JILA experiment gives rise to a rich spatial
separation dynamics between the two condensates \cite{JILA_demix},
which complicates the theoretical study of the relative phase dynamics. 
As a consequence
previous theoretical treatments of the phase decoherence 
processes, dealing essentially
with steady state condensates, as in \cite{Bigelow}, cannot {\it a priori}
be applied to the experimental situation.

A treatment of the phase coherence of two interacting, non
stationary, condensates
can be found in \cite{Villain}, with two important differences as
compared to the present situation of interest: (1) In \cite{Villain}
the condensates are
subject to a continuous coherent coupling of amplitude $\Lambda$;
results
are obtained from a perturbative expansion in powers
of $1/\Lambda$ and cannot be simply extended to the present $\Lambda=0$ case.
(2) In \cite{Villain} all the coupling constants $g_{aa},g_{ab},g_{bb}$
between the two internal atomic states $a$ and $b$
are assumed to be equal.

In this paper we propose a formalism to study the relative 
phase dynamics of interacting and dynamically evolving Bose-Einstein
condensates
initially at zero temperature.

We present the general method in section \ref{sec:gen-method}. It
consists in expanding the initial state on Fock states, and 
in evolving each Fock state in the Hartree-Fock approximation.
We determine the time dependence of the phase collapse for a binary mixture 
of condensates, due to (1) fluctuations in the relative number
of particles between the condensates, intrinsic to the initial state
with well defined relative phase, and (2) fluctuations in the
total number of particles.
In the next two sections 
we apply this general formalism to two limiting cases that can be treated
analytically.

The first case, in section
\ref{sec:breath-together}, considers a particular solution of the 
non-linear Schr\"odinger equations for the condensates wavefunctions;
in this solution the
two condensates remain spatially superimposed as they breathe
in phase, provided that dynamical stability
conditions (that we determine)
are satisfied. We find that phase decoherence can be highly reduced
with respect to non mutually interacting condensates
when the three coupling constants $g_{aa},
g_{ab},g_{bb}$ between atoms in the two internal states $a,b$ are
close to each other. 

In section \ref{sec:closeg's},
we therefore study in a more general case (not restricted to the
breathe-together solution)
the dynamics of the relative phase for a
mixture of condensates for close coupling constants.
Our treatment requires also in this case
the absence of demixing instability, a point that we discuss
in detail.

Finally we discuss the case of the JILA experiment in section \ref{sec:JILA}.
This case, that corresponds to close coupling constants in a regime
of demixing instability, is more difficult to analyze. 
The predicted
phase collapse time depends on the fluctuations of
the total number of particles ; it is on the order
of 0.4 second for Gaussian fluctuations of 8\%.

\section{General method}
\label{sec:gen-method}
In this section, we introduce a gedanken experiment 
to characterize phase coherence
between two condensates: the relevant quantity is the  interference
term $\langle\hat{\psi_b}^\dagger(\vec{r},t)\hat{\psi_a}(\vec{r},t)\rangle$
between the atomic fields of the two condensates $a$ and $b$.
Subsequently we express this interference term in the Hartree-Fock
approximation, assuming an initially well defined relative
phase between the condensates. After a further approximation on the modulus and 
the phase of the condensate wavefunctions, we determine the decay with time
of the interference term due to atomic interactions; we arrive at the
simple results Eq.(\ref{eq:inter_court}) for a fixed total number of particles 
and Eq.(\ref{eq:gaus_court}) for Gaussian fluctuations in the total number of
particles.

\subsection{Considered gedanken experiment}
The experimental procedure we consider to measure the phase
coherence is inspired by recent experiments at JILA 
\cite{JILA_phase}. A condensate is first created in a trap in
some internal atomic state $a$; the corresponding condensate wavefunction
in the zero temperature mean-field approximation is $\phi_0$, 
a stationary solution of the Gross-Pitaevskii
equation:
\begin{equation}
\mu\phi_0 = -{\hbar^2\over 2m} \Delta\phi_0 +
[U_a(\vec{r})+Ng_{aa}|\phi_0|^2]\phi_0.
\label{eq:phi0}
\end{equation}
In this equation $N$ is the number of particles, $g_{aa}$
is the coupling constant between the atoms in the internal state $a$,
related to the scattering length $a_{aa}$ by $g_{aa}=4\pi\hbar^2 a_{aa}
/m$;
$U_a$ is the trapping potential seen by the atoms in $a$ and $\mu$
is the chemical potential. Note that we have normalized $\phi_0$
to unity.

At time $t=0$ a resonant electromagnetic pulse transfers in a coherent way
part of the atoms to a second internal state $b$. The state of
the system is then given in the Hartree-Fock approximation by
\begin{equation}
|\Psi(0)\rangle = [c_a |a,\phi_0\rangle + c_b |b,\phi_0\rangle]^N
\label{eq:init}
\end{equation}
with $|c_a|^2 + |c_b|^2 =1$. As we assume a Rabi coupling between $a$ and $b$
much more intense than $\mu/\hbar$
the atomic interactions have a negligible effect during the transfer
so that the amplitudes $c_{a,b}$ depend only
on the pulse parameters, not on the number $N$ of particles.
In the $N$-particle state Eq.(\ref{eq:init})
the condensate in state $a$ and the condensate
in state $b$ have a well defined relative phase; we therefore call
this state a {\sl phase state}, in analogy with \cite{Leggett}.

The two condensates evolve freely in their trapping potentials
during the time $\tau$. During this evolution we assume that
there is no coherent
coupling between $a$ and $b$ to lock the relative phase
of the condensates; in particular the only considered  interactions between
the particles are elastic, of the type $a+a\rightarrow a+a$ (coupling
constant $g_{aa}>0$),
$a+b\rightarrow a+b$ (coupling constant $g_{ab}>0$),
$b+b\rightarrow b+b$ (coupling constant $g_{bb}>0$).
We therefore expect a collapse of the relative phase for sufficiently
long times, due to atomic interactions.

To test the phase coherence at time $\tau$, a second electromagnetic
pulse is applied to mix the internal states $a$ and $b$. We assume that
this second pulse is a $\pi/2$ pulse, so that the atomic field operators
in the Heisenberg picture 
are transformed according to
\begin{eqnarray}
\hat{\psi}_a(\tau^+) & = & {e^{-i\delta}\over\sqrt{2}} \hat{\psi}_a(\tau^-)
+ {e^{i\delta}\over\sqrt{2}} \hat{\psi}_b(\tau^-) \\
\hat{\psi}_b(\tau^+) & = & -{e^{-i\delta}\over\sqrt{2}} \hat{\psi}_a(\tau^-)
+ {e^{i\delta}\over\sqrt{2}} \hat{\psi}_b(\tau^-),
\end{eqnarray}
$\delta$ being an adjustable phase.
One then measures the mean spatial density $\rho_a$ in the internal state $a$,
averaging over many realizations of the whole experiment:
\begin{equation}
\rho_a = \langle \hat{\psi}^\dagger_a(\tau^+)\hat{\psi}_a(\tau^+)\rangle.
\end{equation}

The signature of a phase coherence between the two condensates
at time $\tau$  is the dependence of the mean density
$\rho_a$ on the adjustable phase $\delta$. More precisely
we define the contrast
\begin{equation}
C ={ \mbox{max}_\delta \rho_a -\mbox{min}_\delta\rho_a
\over
\mbox{max}_\delta\rho_a
+\mbox{min}_\delta\rho_a}
=
{2|\langle\hat{\psi}_b^\dagger(\tau^-)\hat{\psi}_a(\tau^-)\rangle|
\over \sum_{\varepsilon=a,b} 
\langle \hat{\psi}^\dagger_\varepsilon(\tau^-)
\hat{\psi}_\varepsilon(\tau^-)\rangle}
\end{equation}
The contrast involves the interference term 
$\langle\hat{\psi}_b^\dagger(\tau^-)\hat{\psi}_a(\tau^-)\rangle$
which carries the information about the relative phase between the
two condensates.

\subsection{Approximate evolution of an initial phase state}

The time evolution in the phase state
representation is not simple, as an initial phase state
is mapped onto a superposition of phase states. It is
more convenient to introduce Fock states, that is states
with a well defined number of particles in $a$ and in $b$, 
these numbers being preserved by the time evolution.

We therefore expand the initial phase state over the Fock states:
\begin{equation}
|\Psi(0)\rangle = \sum_{N_a=0}^{N} \left({N!\over N_a! N_b!}\right)^{1/2}
c_a^{N_a}c_b^{N_b}|N_a:\phi_0,N_b:\phi_0\rangle
\end{equation}
where we set $N_b=N-N_a$.

By calculating the evolution of each Fock state in the simplest Hartree-Fock
approximation, we get the following mapping:
\begin{equation}
|N_a:\phi_0,N_b:\phi_0\rangle \rightarrow 
e^{-iA(N_a,N_b;t)/\hbar}
|N_a:\phi_a(N_a,N_b;t),N_b:\phi_b(N_a,N_b;t)\rangle
\label{eq:evol_Fock}
\end{equation}
where the condensates wavefunctions evolve according to the 
coupled Gross-Pitaevskii equations:
\begin{equation}
i\hbar\partial_t\phi_\varepsilon = \left[-{\hbar^2\over 2m}\Delta + 
U_\varepsilon(\vec{r})
+N_\varepsilon g_{\varepsilon\varepsilon}|\phi_\varepsilon|^2
+N_{\varepsilon'} g_{\varepsilon\varepsilon'}|\phi_{\varepsilon'}|^2\right]
\phi_\varepsilon
\label{eq:gpe}
\end{equation}
(where $\varepsilon'\neq\varepsilon$)
with the initial conditions
\begin{equation}
\phi_a(0)=\phi_b(0)=\phi_0
\label{eq:init_cond}
\end{equation}
and where 
the time dependent phase factor $A$ solves:
\begin{equation}
{d\over dt} A(N_a,N_b;t) = -{1\over 2} N_a^2 g_{aa}\int
d\vec{r}\;|\phi_a|^4
-{1\over 2} N_b^2 g_{bb}\int d\vec{r}\;|\phi_b|^4
-N_aN_b g_{ab} \int d\vec{r}\;|\phi_a|^2|\phi_b|^2.
\label{eq:apoint}
\end{equation}
Equation (\ref{eq:apoint}) is derived in Appendix \ref{app:A}.
Physically $dA/dt$ is simply the opposite of the mean interaction
energy between the particles in the Fock state.
In the case where the Fock state is a steady state,
the need for the phase factor $A$ additional to the Gross-Pitaevskii
equation is obvious;
the exact phase factor is indeed $e^{-iEt/\hbar}$, where $E$ is the
energy of the Fock state, whereas the phase factor obtained 
from the Gross-Pitaevskii evolution is $e^{-i(N_a\mu_a+N_b\mu_b)t/\hbar}$,
where $\mu_{a,b}$ is the chemical potential in $a,b$.

Using the evolution of the Fock states, and other approximations
valid in the limit of large numbers of particles (as detailed
in the Appendix \ref{app:interf}) we obtain for
the interference term between the condensates with a well-defined total number
$N$ of particles:
\begin{equation}
\langle \hat{\psi}_b^\dagger \hat{\psi_a}\rangle _{N} 
= c_a c_b^* \sum_{N_a=1}^{N} {N!\over (N_a-1)!N_b!}
|c_a|^{2(N_a-1)}|c_b|^{2N_b}
\phi_a(N_a,N_b)\phi_b^*(N_a-1,N_b+1)
\label{eq:complique}
\end{equation}
where $N_b=N-N_a$.
The exact computation of this sum remains a formidable task, since
it involves in principle the solution of $N$ different sets of two coupled
Gross-Pitaevskii equations. We introduce some simplifying approximations
in the next subsection.

\subsection{Phase collapse for a mixture}
In the present experiments the total number of particles
fluctuates from one realization to the other, so that Eq.(\ref{eq:complique})
has to be averaged over $N$.
We assume that the fluctuations of the total number of particles have
a standard deviation $\Delta N$ much smaller than the mean total particle
number $\bar{N}$.
As the distributions of the number of particles
in $a$ and $b$ in a phase state have also a width much
smaller than $\bar{N}$ (typically on the order of $\bar{N}^{1/2}$)
we can assume than the number of particles in $a$ and in $b$
are very close to their average values
$\bar{N}_\varepsilon = |c_\varepsilon|^2\bar{N}$.
We now take advantage of this property to simplify Eq.(\ref{eq:complique}).

We split the condensate wavefunctions in a modulus and a phase 
$\theta_\varepsilon$;
we assume that the variation of the modulus can be neglected 
over the distribution of
$N_{a,b}$, and that the variation of the phase 
can be approximated by a linear expansion
around $\bar{N_\varepsilon}$. 
We thus get the approximate form
for the condensate wavefunctions:
\begin{equation}
\phi_\varepsilon(N_a,N_b) \simeq \bar{\phi_\varepsilon}
\exp\left[i\sum_{\varepsilon'=a,b}(N_{\varepsilon'}-\bar{N_{\varepsilon'}})
(\partial_{N_{\varepsilon'}}\theta_{\varepsilon})(\bar{N_a},\bar{N_b})\right]
\end{equation}
where $\bar{\phi_\varepsilon} = \phi_\varepsilon(N_a=\bar{N_a},
N_b=\bar{N_b})$.

To this level of approximation the mean densities in the internal
states $a,b$ are simply given by
\begin{equation}
\langle\hat{\psi}_\varepsilon^\dagger\hat{\psi}_\varepsilon\rangle_N
\simeq \bar{N_\varepsilon}|\bar{\phi_\varepsilon}|^2
\end{equation}
whereas the interference term
between the condensates is:
\begin{eqnarray}
\langle \hat{\psi}_b^\dagger \hat{\psi_a}\rangle _{N}
&\simeq& \bar{N}c_ac_b^*\bar{\phi_a}\bar{\phi_b}^*
\exp\left\{i[(N-\bar{N})\chi_s-\bar{N}(|c_a|^2-|c_b|^2)\chi_d]
\right\}e^{i\chi_0} 
\nonumber\\
&&\left[|c_a|^2 e^{i\chi_d} + |c_b|^2 e^{-i\chi_d}\right]^{N-1}.
\label{eq:inter_N}
\end{eqnarray}
In this last expression
we have introduced the time and position dependent quantities
\begin{eqnarray}
\chi_s &=& {1\over 2} \left[\left(\partial_{N_a}+\partial_{N_b}\right)
(\theta_a-\theta_b)\right](\bar{N_a},\bar{N_b}) 
\label{eq:chi_s} \\
\chi_d &=& {1\over 2} \left[\left(\partial_{N_a}-\partial_{N_b}\right)
(\theta_a-\theta_b)\right](\bar{N_a},\bar{N_b}).
\label{eq:chi_d}
\end{eqnarray}
The phase 
$\chi_0 = {1\over 2} (\partial_{N_a}-\partial_{N_b})(\theta_a
+\theta_b)(\bar{N_a},\bar{N_b})$ in Eq.(\ref{eq:inter_N})
is less important as contrarily
to $\chi_{s,d}$ it is not multiplied by $N$.
At time $t=0$ all the $\chi$'s
vanish.
In the large $N$ limit, the $\chi$'s are expected
to be on the order of $\bar{\mu} t/\hbar\bar{N}$.

The factor responsible for the collapse of the contrast at a 
fixed value of $N$ is the second line of Eq.(\ref{eq:inter_N}),
the factors in the first line being of modulus one.
As $N$ is large a small variation of $\chi_d$ from its initial value
$\chi_d(t=0)=0$ is sufficient to destroy the interference term.
Over the range of the collapse we can therefore expand the
exponential of $\pm i\chi_d$ to second order in $\chi_d$, obtaining:
\begin{equation}
\langle \hat{\psi}_b^\dagger \hat{\psi_a}\rangle _{N}
\simeq \bar{N}c_ac_b^*\bar{\phi_a}\bar{\phi_b}^*
\exp\left\{i[(N-\bar{N})[\chi_s+(|c_a|^2-|c_b|^2)\chi_d]\right\}\ 
\exp\left[-2N\chi_d^2|c_a|^2|c_b|^2\right].
\label{eq:inter_court}
\end{equation}
The second exponential factor 
in this expression allows to determine the collapse
time $t_c^{\mbox{\scriptsize fix}}$ for a fixed number of particles, through the condition
\begin{equation}
4N|c_a|^2|c_b|^2\chi_d^2(t_c^{\mbox{\scriptsize fix}}) \simeq 1
\label{eq:deftc}
\end{equation}
such that the modulus of the interference term is reduced
by a factor $e^{-1/2}$ from its initial value.
The first exponential factor in Eq.(\ref{eq:inter_court})
accounts for the phase
difference of the interference term for $N$ particles
and $\bar{N}$ particles, as shown by the identity:
\begin{equation}
\chi_s+(|c_a|^2-|c_b|^2)\chi_d = {d\over dN}
\left[(\theta_a-\theta_b)(N|c_a|^2,N|c_b|^2)\right]_{N=\bar{N}}.
\label{eq:magique}
\end{equation}
This phase factor can also be understood as a consequence of a
drift of the relative phase between two condensates at
a velocity $v(N)$ depending on the total number of particles:
\begin{equation}
v(N) = {\bar{\mu_b}-\bar{\mu_a}\over \hbar} +
(N-\bar{N}) \left[\chi_s+(|c_a|^2-|c_b|^2)\chi_d\right].
\label{eq:vdrift}
\end{equation}
As we shall see in  the next subsection
fluctuations in the total number of particles $N$ result in
fluctuations of this phase factor, providing an additional source
of smearing of the phase, as already emphasized in \cite{resur}.

\subsection{Effect of fluctuations in the total number of particles}
The effect on the phase collapse
of fluctuations in the total number of particles
is obtained by averaging Eq.(\ref{eq:inter_court}) over the
probability distribution of $N$.
To be specific we assume a Gaussian distribution for $N$.
The average can be calculated by replacing the discrete sum
over $N$ by an integral; we neglect a term proportional to 
$(\Delta N\chi_d^2)^2$ scaling as $(\Delta N/\bar{N})^2$ at the
collapse time $t_c^{\mbox{\scriptsize fix}}$; the resulting modulus of the
interference term reads:
\begin{equation}
|\langle \hat{\psi}_b^\dagger \hat{\psi_a}\rangle^{\mbox{\scriptsize Gauss}}|
\simeq  \bar{N}|c_ac_b^*\bar{\phi_a}\bar{\phi_b}^*|
\exp\left\{
-{1\over 2}(\Delta N)^2
\left[{d\over dN}(\theta_a-\theta_b)\right]^2_{N=\bar{N}} 
\right\}
\ 
\exp\left[-2\bar{N}\chi_d^2|c_a|^2|c_b|^2\right].
\label{eq:gaus_court}
\end{equation}
The first exponential factor in this expression represents
the damping of the interference term due to fluctuations 
in the total number of particles; the second exponential factor,
already present in Eq.(\ref{eq:inter_court}), gives the damping 
due to fluctuations in the relative number of particles 
between $a$ and $b$, as can be seen in Eq.(\ref{eq:chi_d}).

\subsection{The steady state case and
comparison with previous treatments} 
\label{subsec:sta}

Our treatment can be easily adapted to the case of two initially
different condensate wavefunctions $\phi_a(t=0)$ and
$\phi_b(t=0)$.
In the particular case of 
condensates in stationary states, 
the formulas for the interference
term $\langle\hat{\psi_b}^\dagger\hat{\psi_a}\rangle$
remain the same, and 
one has $\theta_\varepsilon=
-\mu_\varepsilon(N_a,N_b)t/\hbar$.
We can give in this case the explicit expression
for the collapse time $t_c^{\mbox{\scriptsize fix}}$ 
defined in Eq.(\ref{eq:deftc}), assuming
a fixed total number of atoms $N=\bar{N}$:
\begin{equation}
t_c^{\mbox{\scriptsize fix}} = \hbar \left[\bar{N}^{1/2}|c_a c_b|
|(\partial_{N_a}-\partial_{N_b})(\mu_a-\mu_b)|
\right]^{-1}.
\label{eq:geniale}
\end{equation}
For the particular case of non mutually interacting steady state condensates
$\mu_\varepsilon$ depends on $N_\varepsilon$ only, so that
the partial derivatives in the denominator of Eq.(\ref{eq:geniale})
reduce to $d\mu_a/dN_a + d\mu_b/dN_b$, and we recover
the results of \cite{Jean,resur}.

From Eq.(\ref{eq:geniale})  we see that what matters physically
is the change in the {\it difference} between the chemical
potentials of the two condensates when one transfers one particle
from one condensate to the other. For this reason 
the case of mutually interacting condensates with close coupling
constants can lead to much larger $t_c$'s as compared to
the case of non-mutually interacting condensates.
For example, in the case of the JILA experiment \cite{JILA_phase},
assuming that the condensates are in steady state,
one finds $t_c^{\mbox{\scriptsize fix}} \simeq 3.1$ s; by ignoring the interaction between
the condensates (setting by hand $g_{ab}=0$) one obtains the much
shorter time $ \simeq 0.25$ s. The JILA case is analyzed in more
detail in our section \ref{sec:JILA}.

A similar prediction on the reduction of decoherence due to
mutual interactions between the two condensates, in
trapping potentials with different curvatures, was obtained numerically
in \cite{Bigelow}.

The treatment in \cite{Maciek} considers the absolute
phase dynamics of a single condensate
(in our formalism $c_b=0$) in a coherent state.
When the condensate wavefunction is stationary one has simply 
$\theta_a = -\mu_a t/\hbar$.
From Eq.(\ref{eq:gaus_court}) with $\Delta N=\bar{N}^{1/2}$
(as the coherent state has a Poisson distribution for $N$)
we then find that the phase of the condensate order parameter
is damped as $\exp[-\bar{N}(d\mu_a/dN)^2t^2/2\hbar^2]$ as in \cite{Maciek}.

\section{Application to the breathe-together solution}
\label{sec:breath-together}

In this section we consider a particular solution of the coupled
Gross-Pitaevskii equations for which an approximate scaling
solution is available when the chemical potential is much 
larger than the energy spacing between trap levels, the so-called
Thomas-Fermi regime.
We first give the set of parameters for which this solution, that we
call the breathe-together solution, exists. We then linearize the
Gross-Pitaevskii equations around this solution to determine 
its stability with respect to demixing and to obtain the 
phase coherence dynamics.

\subsection{Description of the breathe-together solution}
We now determine the set of parameters such that the coupled
Gross-Pitaevskii equations Eq.(\ref{eq:gpe}) for
\begin{equation}
N_\varepsilon = \bar{N_\varepsilon}\equiv \bar{N} |c_\varepsilon|^2
\end{equation}
have a solution
with  $\bar{\phi_a}(\vec{r},t)=\bar{\phi_b}(\vec{r},t) 
\equiv\bar{\phi}(\vec{r},t)$.
The general condition is that the effective potential, that is
the trapping potential plus the mean field potential, seen by the atoms
in $a$ and in $b$ should be the same. This condition is satisfied when:
\begin{eqnarray}
U_a(\vec{r}) = U_b(\vec{r}) &\equiv& U(\vec{r}) \\
\bar{N_a} g_{aa} + \bar{N_b} g_{ab} = \bar{N_b} g_{bb} + {\bar N_a} 
g_{ab} &\equiv& \bar{N} g.
\label{eq:angle}
\end{eqnarray}

The resulting Gross-Pitaevskii equation for the condensate
wavefunction $\bar{\phi}$ common to $a$ and $b$ is then:
\begin{equation}
i\hbar\partial_t\bar{\phi} =\left[-{\hbar^2\over 2m}\Delta + U(\vec{r})
+\bar{N}g |\bar{\phi}|^2\right]\bar{\phi}
\end{equation}
with the initial condition $\bar{\phi}(\vec{r},0)=\phi_0[N=\bar{N}](\vec{r})
\equiv\bar{\phi_0}$,
where $\phi_0$ is defined in Eq.(\ref{eq:phi0}).

By rewriting Eq.(\ref{eq:angle}) as
$\bar{N_a}/\bar{N_b}=(g_{bb}-g_{ab})/(g_{aa}-g_{ab})$ we see that this equality
can be satisfied by choosing properly the mixing angle between
$a$ and $b$ provided that
\begin{equation}
g_{ab} < g_{aa},g_{bb} \ \ \ \  \mbox{or} \ \ \ \  \ \ g_{ab} > g_{aa},g_{bb}.
\end{equation}
As we shall see below, only the first case is relevant here,
since the second case corresponds
to an unstable solution with respect to demixing between 
$a$ and $b$.

\subsection{Linearization around the breathe-together solution}
The strategy to obtain the quantities $\chi_{s,d}$ relevant
for the phase dynamics is to calculate in the linear 
approximation the deviations
$\delta\phi_\varepsilon$
between the breathe-together solution $\bar{\phi}$ and neighboring solutions
$\phi_{\varepsilon}$ for $N_\varepsilon$ slightly different
from $\bar{N_\varepsilon}$:
\begin{equation}
\delta\phi_\varepsilon \equiv \phi_\varepsilon(\bar{N_a}+\delta N_a,\bar{N_b}
+\delta N_b)-\phi_\varepsilon(\bar{N_a},\bar{N_b}).
\label{eq:def_delta_phi}
\end{equation}
From the definitions Eq.(\ref{eq:chi_s}) and Eq.(\ref{eq:chi_d})
one indeed realizes that  in the limit of small $\delta N_a$:
\begin{eqnarray}
\chi_s &=&\left[{\delta\theta_a-\delta\theta_b\over 2\delta N_a}
\right]_{\delta N_b=\delta N_a} \label{eq:chi1}\\
\chi_d &=&\left[{\delta\theta_a-\delta\theta_b\over 2\delta N_a}
\right]_{\delta N_b=-\delta N_a} 
\label{eq:chi2}
\end{eqnarray}
where $\delta\theta_{a,b}$ are the deviations of the phase
of the neighboring solutions $\phi_{\varepsilon}$ from the phase
of the breathe-together solution:
\begin{equation}
\delta\theta_a-\delta\theta_b = \mbox{Im} \left[{\delta\phi_a\over\bar{\phi}}
-{\delta\phi_b\over\bar{\phi}}\right].
\label{eq:phase_rel_def}
\end{equation}

It turns out that homogeneous rather than inhomogeneous linear equations
can be obtained for the deviations $\delta\phi_\epsilon$
by introducing the quantities:
\begin{equation}
\delta\varphi_\varepsilon
\equiv{\delta[\sqrt{N_\varepsilon}\phi_\varepsilon]
\over \sqrt{\bar{N_\varepsilon}}}= {\delta N_\varepsilon\over 2
\bar{N_\varepsilon}}
\bar{\phi} + \delta\phi_\varepsilon.
\label{eq:def_varphi}
\end{equation}
Furthermore a partial decoupling occurs for the linear
combinations
\begin{eqnarray}
\delta\varphi_s &\equiv &\delta\varphi_a + \delta\varphi_b \\
\delta\varphi_d &\equiv &\delta\varphi_a - \delta\varphi_b.
\end{eqnarray}
The sum $\delta\varphi_s$ obeys a linear equation
involving $\delta\varphi_d$ as a source term:
\begin{eqnarray}
i\hbar\partial_t \delta\varphi_s& =&\left[-{\hbar^2\over 2m}\Delta +U +
2\bar{N}g |\bar{\phi}|^2\right]\delta\varphi_s\nonumber\\
& +& \bar{N}g\bar{\phi}^2\delta\varphi_s^*
+(\bar{N_a}g_{aa}-\bar{N_b}g_{bb})
(|\bar{\phi}|^2\delta\varphi_d + \bar{\phi}^2\delta\varphi_d^*).
\end{eqnarray}
The part of this equation involving $\delta\varphi_s$ is identical
to the one obtained for a single condensate with $\bar{N}$ particles
and a coupling constant $g$. The corresponding modes have minimal
frequencies on the order of the trap frequency $\omega$
for an isotropic harmonic trap \cite{Stringari}.

The difference $\delta\varphi_d$ obeys the closed equation:
\begin{equation}
i\hbar\partial_t \delta\varphi_d = \left[-{\hbar^2\over 2m}\Delta +U +
\bar{N}g |\bar{\phi}|^2\right]\delta\varphi_d +  {\bar{N_a}\bar{N_b}
\over \bar{N}} (g_{aa}+g_{bb}-2g_{ab})
(|\bar{\phi}|^2\delta\varphi_d + \bar{\phi}^2\delta\varphi_d^*)
\end{equation}
where we have used the identity:
\begin{equation}
\bar{N_b}(g_{bb}-g_{ab}) = \bar{N_a}(g_{aa}-g_{ab}) = {\bar{N_a}\bar{N_b}
\over \bar{N}} (g_{aa}+g_{bb}-2g_{ab}).
\end{equation}
As shown in \cite{Nous} 
minimal eigenfrequencies of this equation can be much smaller than
$\omega$; e.g.\ when all the coupling constants are equal, the
minimal eigenfrequencies in a harmonic isotropic trap
of frequency $\omega$ scale as $\hbar\omega^2/\mu\ll\omega$ in the Thomas-Fermi
limit.

For the derivation of the $\chi$'s 
it is sufficient to calculate
$\delta\varphi_d$. The relative phase between the two
condensates for the considered neighboring solution with
$N_\varepsilon=\bar{N_\varepsilon}+\delta N_\varepsilon$ particles
in the state $\varepsilon$ is in fact given by:
\begin{equation}
\delta\theta_a-\delta\theta_b 
= {1\over 2i}\left[ {\delta\varphi_d\over \bar{\phi}} -{\delta\varphi_d^*\over
\bar{\phi}^*}\right].
\label{eq:rel_phase}
\end{equation}
as can be checked from the definition Eq.(\ref{eq:phase_rel_def}).

\subsection{Approximate equations of evolution in the Thomas-Fermi limit}
In the remaining part of this section we assume an isotropic harmonic 
trapping potential $U(\vec{r})=m\omega^2 r^2/2$ and we restrict to the
Thomas-Fermi limit $\mu\gg\hbar\omega$. 

In the Thomas-Fermi limit 
it is known \cite{Kagan,Yvan} that most of the time dependence
of the wavefunction $\bar{\phi}$ can be absorbed by a time dependent gauge and
scaling transform; here we apply this transform to
both $\bar{\phi}$ and $\delta\varphi_d$:
\begin{eqnarray}
\bar{\phi}(\vec{r},t) & \equiv & {e^{-i\eta(t)}\over\lambda^{3/2}(t)}
e^{imr^2\dot{\lambda}(t)/2\hbar\lambda(t)}
\tilde{\bar{\phi}}(\vec{r}/\lambda(t),t) 
\label{eq:gau1}\\
\delta\varphi_d(\vec{r},t) & = & {e^{-i\eta(t)}\over\lambda^{3/2}(t)}
e^{imr^2\dot{\lambda}(t)/2\hbar\lambda(t)}
\tilde{\delta\varphi_d}(\vec{r}/\lambda(t),t).
\label{eq:gau2}
\end{eqnarray}
The scaling factor $\lambda(t)$ solves the second order Newton-type
differential equation
\begin{equation}
\ddot{\lambda}={g\over g_{aa}}{\omega^2\over\lambda^4} -\omega^2\lambda
\label{eq:ev_lambda}
\end{equation}
with the initial condition $\lambda(0)=1,\dot{\lambda}(0)=0$.
The \lq\lq force" seen by $\lambda$ in Eq.(\ref{eq:ev_lambda})
derives from the sum of an expelling $1/\lambda^3$ potential 
due to repulsive interactions between atoms and an attractive
$\lambda^2$ potential due to the harmonic confinement of the atoms.
It leads to periodic oscillations of $\lambda$, that is
to a periodic breathing of the condensates.
We have also introduced a phase factor involving the time dependent
function $\eta$ such that
$\dot{\eta} = \bar{\mu} g/(g_{aa}\lambda^3\hbar)$.

In the Appendix \ref{app:TF} we derive approximate evolution equations
for $\tilde{\bar{\phi}}$ 
and $\tilde{\delta\varphi_d}$; we give here only the result.
To lowest order in the Thomas-Fermi approximation
$\tilde{\bar{\phi}}$ does not evolve and can be approximated by the Thomas-Fermi
approximation for $\bar{\phi_0}$:
\begin{equation}
\tilde{\bar{\phi}}(\vec{r},t) \simeq
\bar{\phi_0}(\vec{r}) \simeq \left({15\over 8\pi R_0^3}\right)^{1/2}
\left[1-{r^2\over R_0^2}\right]^{1/2}
\label{eq:phi_TF}
\end{equation}
with a Thomas-Fermi radius $R_0=\sqrt{2\bar{\mu}/m\omega^2}$.
The approximate evolution for $\tilde{\delta\varphi_d}$ is conveniently
expressed in terms of the real function $\alpha$ and the purely
imaginary function $\beta$:
\begin{eqnarray}
\alpha &=& \tilde{\bar{\phi}}^*\tilde{\delta\varphi_d}+
\tilde{\bar{\phi}}\tilde{\delta\varphi_d}^*
\label{eq:def_alpha}\\
\beta &=& {1\over 2}\left[ {\tilde{\delta\varphi_d}\over\tilde{\bar{\phi}}}
-{\tilde{\delta\varphi_d}^*\over\tilde{\bar{\phi}}^*}\right].
\label{eq:def_beta}
\end{eqnarray}
These functions have a clear physical meaning. The first one $\alpha$
corresponds to the deviation $\delta\rho_a/\bar{N_a}-\delta\rho_b/\bar{N_b}$
written in the rescaled frame, $\delta\rho_\varepsilon$ being the deviation
of spatial density in the condensate $\varepsilon$ from the breathe-together
solution.
Apart from a factor $i$ the second function $\beta$ is the deviation
of the relative phase Eq.(\ref{eq:rel_phase}) written in the rescaled
frame:
\begin{equation}
(\delta\theta_a-\delta\theta_b)(\vec{r},t)=
-i\beta(\vec{r}/\lambda,t).
\label{eq:sens_beta}
\end{equation}

The equations of evolution for $\alpha,\beta$ are:
\begin{equation}
i\hbar\partial_t 
\left( \begin{array}{c} \alpha \\ \beta \end{array} \right) =
L(t)
\left( \begin{array}{c} \alpha \\ \beta \end{array} \right)
\label{eq:evol_ab}
\end{equation}
where the operator $L(t)$ in the Thomas-Fermi approximation
reads:
\begin{equation}
L(t)=
\left(\begin{array}{cc}
0 &  -{\hbar^2\over m\lambda^2} 
\; \mbox{div}\left[\bar{\phi_0}^2\;\vec{\mbox{grad}}(\cdot)\right] \\
{1\over\lambda^3} {\bar{N_a}\bar{N_b}
\over \bar{N}} (g_{aa}+g_{bb}-2g_{ab}) & 0 \end{array}\right).
\label{eq:op_ab}
\end{equation}
The initial conditions for $\alpha,\beta$ at time $t=0$
obtained from Eq.(\ref{eq:def_varphi}) and Eq.(\ref{eq:init_cond}) are:
\begin{eqnarray}
\label{eq:init_alpha}
\alpha(0) &=& \left({\delta N_a\over \bar{N_a}}
-{\delta N_b\over \bar{N_b}}\right)
\bar{\phi_0}^2 \\
\beta(0) &=& 0.
\label{eq:init_beta}
\end{eqnarray}

\subsection{Solution of the Thomas-Fermi evolution equations: stability
against demixing}
The strategy to determine the time evolution of $\alpha,\beta$
is (1) to expand the vector $(\alpha(0),\beta(0))$ on eigenmodes
of the operator $L(0)$, and (2) to calculate
the time evolution of each eigenmode.

\subsubsection{Expansion on modes of $L(0)$}
Consider an eigenvector $(\alpha,\beta)$ of the operator $L(0)$
with the eigenvalue $\hbar\Omega$. For $\Omega\neq 0$ one
can express the component $\beta$ as function of $\alpha$:
\begin{equation}
\beta = {\alpha\over\hbar\Omega}
 {\bar{N_a}\bar{N_b}
\over \bar{N}} (g_{aa}+g_{bb}-2g_{ab})
\label{eq:elimine_beta}
\end{equation}
and obtain the eigenvalue problem for $\alpha$:
\begin{equation}
\Omega^2\alpha = \left({\bar{N_a}\bar{N_b}
\over \bar{N}^2} {(g_{aa}+g_{bb}-2g_{ab})\over g_{aa}}\right)S[\alpha]
\label{eq:eigen}
\end{equation}
where we have introduced the Stringari operator:
\begin{equation}
S[\alpha] \equiv -{\bar{N}g_{aa}\over m}\;\mbox{div}[\bar{\phi_0}^2\;
\vec{\mbox{grad}}\;\alpha].
\label{eq:stringari}
\end{equation}
This operator has been studied in \cite{Stringari}.
It is an Hermitian and positive operator, with a
spectrum $q\omega^2$, $q$ non negative integer;
$q$ is given by
\begin{equation}
q=2n^2+2nl+3n+l
\end{equation}
as function of the radial quantum number $n$ and the angular momentum
$l$.
This allows the determination of the eigenfrequencies
$\Omega$:
\begin{equation}
\Omega_q = \pm\left({\bar{N_a}\bar{N_b}
\over \bar{N}^2} {(g_{aa}+g_{bb}-2g_{ab})\over g_{aa}}\right)^{1/2}q^{1/2}\omega,
\end{equation}
with $q>0$ as we have assumed $\Omega\neq 0$.
The case of a vanishing $\Omega$ corresponds to the 
zero energy mode $\alpha_0=0,\beta_0=1$ of the operator $L(0)$,
as it can be checked from a direct calculation.

All the eigenmodes of $L(0)$ have been identified. 
They do not form a complete family of vectors however.
The vector $(\alpha=1,\beta=0)$ cannot be expanded on the eigenmodes of $L(0)$. 
Its first component $\alpha$ is indeed in the kernel of the operator
$S$ (as $S[\alpha]=0$) whereas none of the $\alpha_q$ is in the kernel
of $S$ ($S[\alpha_q]=q \omega^2 \alpha_q$ is not identically zero)
except when $\alpha_q$ is identically zero (for $q=0$). 
The family of eigenvectors of $L(0)$ completed by
the additional vector $(\alpha=1,\beta=0)$ forms a basis. 
The additional vector is called an anomalous mode, and we set
$\alpha_{\mbox{\scriptsize anom}} = 1$, $\beta_{\mbox{\scriptsize anom}} = 0$;
the action of $L(0)$
on the anomalous mode gives the zero energy mode
times the constant factor $\bar{N_a}(g_{aa}-g_{ab})$
\cite{anomal}.

The mode functions $\alpha_q$ of the operator $S$ are given in
\cite{Stringari}. It turns out that in the expansion of the initial conditions
for $\alpha,\beta$ Eq.(\ref{eq:init_alpha},\ref{eq:init_beta}), 
only the isotropic eigenmodes of $L(0)$ with $q=5$
and the anomalous mode are involved:
\begin{equation}
\left(\begin{array}{c}\alpha(0) \\ \beta(0)\end{array}\right)=
C_5\left[
\left(\begin{array}{c}\alpha_{q=5}\\ \beta_{q=5}\end{array}\right)+
\left(\begin{array}{c}\alpha_{q=5}\\ -\beta_{q=5}\end{array}\right)
\right]
+C_{\mbox{\scriptsize anom}}
\left(\begin{array}{c} 1 \\ 0 \end{array}\right).
\label{eq:devt_t=0}
\end{equation}
The isotropic eigenmode of $S$ with $q=5$, the so-called breathing
mode, reads
\begin{equation}
\alpha_{q=5}(\vec{r}) = \left[{r^2\over R_0^2} -{3\over 5}\right].
\label{eq:mode5}
\end{equation}
By Eq.(\ref{eq:elimine_beta}) we have
$\beta_{q=5} = \alpha_{q=5} \bar{N_a}\bar{N_b}
(g_{aa}+g_{bb}-2g_{ab})/\bar{N}\hbar
\Omega_{q=5}$.
For the coefficients of the modal expansion 
of $(\alpha(0),\beta(0))$, we obtain
\begin{eqnarray}
C_{\mbox{\scriptsize anom}} &=& 
{3\over 4\pi R_0^3}
\left({\delta N_a\over \bar{N_a}} -{\delta N_b\over \bar{N_b}}\right)
\label{eq:coef} \\
C_{5} &=& -{5\over 4} C_{\mbox{\scriptsize anom}}.
\end{eqnarray}

\subsubsection{Evolution of the modes and stability against demixing}
As a second step we determine the time evolution of the modes
of the operator $L(0)$. If we consider an eigenstate $(\alpha_q(0),\beta_q(0))$
of $L(0)$ with the eigenenergy $\hbar\Omega_q$ and evolve it
according to Eq.(\ref{eq:evol_ab}), we find  that the evolution
reduces to multiplication by purely time dependent factors $A_q(t),B_q(t)$:
\begin{eqnarray}
\alpha_q(\vec{r},t) &=& A_q(t) \alpha(\vec{r},0) \\
\beta_q(\vec{r},t) &=& B_q(t) \beta(\vec{r},0) 
\end{eqnarray}
where the factors satisfy the differential equations:
\begin{eqnarray}
i\dot{A_q} &=& {\Omega_q\over\lambda^2} B_q \\
i\dot{B_q} &=& {\Omega_q\over\lambda^3} A_q
\label{eq:syst}
\end{eqnarray}
with the initial conditions $A_q(0) = B_q(0) =1$.
Note that the zero energy eigenmode does not evolve, as $\Omega_q=0$.
The anomalous mode has to be integrated separately, leading to
\begin{eqnarray}
\alpha_{\mbox{\scriptsize anom}} (\vec{r},t) &=& 1 \\
\beta_{\mbox{\scriptsize anom}} (\vec{r},t) &=& {\bar{N_a}\bar{N_b}
\over \bar{N}} {(g_{aa}+g_{bb}-2g_{ab}) \over i\hbar} 
\int_0^t {dt'\over\lambda^3(t')}.
\label{eq:ev_anom}
\end{eqnarray}

We are now able to address the problem of dynamical stability 
of the breathe-together solution. Dynamical stability requires
that any small deviation of the $\phi_\varepsilon$'s
from the breathe-together solution $\bar{\phi}$ should not grow exponentially
with time. Here an exponential growth of $\alpha$
may correspond to a demixing of the two condensates $a$
and $b$.

A first case of instability occurs when $g_{ab}>g_{aa},g_{bb}$.
In this case the eigenfrequencies $\Omega_q$ are purely imaginary
and $A_q,B_q$ diverge exponentially with time \cite{demons}.
We have checked by a numerical integration of the Gross-Pitaevskii
equations with spherical symmetry
that the spatial distribution then acquires a structure
of alternating shells of $a$ atoms and $b$ atoms
(see Fig.\ref{fig:stab}).

We suppose from now on that $g_{ab}<g_{aa},g_{bb}$.
Instability may still occur in this case due to the periodic time dependence
of the coefficients in the system Eq.(\ref{eq:syst}), as shown in
\cite{Ralph}. We have studied in more detail the stability of the mode
$q=5$, which is the one populated initially
(see Eq.(\ref{eq:devt_t=0})); we have 
found non-zero instability exponents $\sigma$
($C_5(t)\sim e^{\sigma t}$) in a very limited region of 
the plane $(g_{ab}/g_{aa},g_{bb}/g_{aa})$, with very small exponents
($\sigma<3\times 10^{-2}\omega$).
A direct numerical integration
of the Gross-Pitaevskii equations
did not show any demixing of $a$ and $b$ even at times $\gg \sigma^{-1}$
\cite{Alice_est_contre}.
This suggests that
the finite instability exponent is an artifact of the Thomas-Fermi
approximation.

We assume in what follows the dynamical stability of
the breathe-together solution.

\begin{figure}
\caption{Modulus squared of the condensate wavefunctions 
$|\phi^2_{a,b}|(N_a,N_b)$ 
as function of the distance
$r$ to the trap center at a time $\omega t\simeq 29.5$,
from a numerical 
solution of the coupled Gross-Pitaevskii
equations in the case of a dynamically unstable breathe-together
solution. We have taken $g_{bb}/g_{aa}=1.2$ and $g_{ab}/g_{aa}=1.5$.
We have applied a deviation $\delta N_a = -\delta N_b=-0.05 \bar{N_a}$
from the exact breathe-together condition. The chemical potential 
is $\bar{\mu}=28.9\hbar\omega$. The curve in solid line corresponds to
$\phi_a$, the dotted curve corresponds to $\phi_b$.
\label{fig:stab}}
\end{figure}

\subsection{Phase dynamics}

In order to calculate the functions $\chi_d,\chi_s$ relevant
for the relative phase dynamics,
we calculate the evolution of the deviation $\delta\varphi_d$
due to a small change in $N_a,N_b$ with respect to $\bar{N_a},
\bar{N_b}$, that is we evolve the initial state
Eq.(\ref{eq:devt_t=0}) according to the results of the previous
subsection.

As we assume dynamical stability of the breathe-together solution,
the modes with $q=5$ perform only oscillations in time \cite{proviso}.
The relevant contribution for the phase dynamics therefore
comes from the anomalous mode, which from Eq.(\ref{eq:ev_anom})
has a $\beta$ diverging linearly 
with time. Assuming $\beta(\vec{r},t) \sim C_{\mbox{\scriptsize anom}}
\beta_{\mbox{\scriptsize anom}} (t)$ and using Eqs.(\ref{eq:coef}),
(\ref{eq:sens_beta}) we obtain:
\begin{equation}
(\delta\theta_a -\delta\theta_b)(\vec{r},t)  \sim 
-{2\bar{\mu}\over 5}{\bar{N_a}\bar{N_b}
\over \bar{N}^2} {(g_{aa}+g_{bb}-2g_{ab})\over g_{aa}}
\left({\delta N_a\over \bar{N_a}} -{\delta N_b\over \bar{N_b}}\right)
\int_0^t {dt' \over\lambda^3(t')}.
\end{equation}
We specialize this formula with $\delta N_b=\pm\delta N_a$
and we get from 
Eq.(\ref{eq:chi1}), Eq.(\ref{eq:chi2}):
\begin{eqnarray}
\chi_d &\sim & -{1\over 2\hbar} \left({d\mu\over dN}\right)_{N=\bar{N}} 
{g_{aa}+g_{bb} -2 g_{ab}\over g_{aa}}
\int_0^t {dt' \over\lambda^3(t')} \\
\chi_s &\sim & (|c_b|^2-|c_a|^2) \chi_d.
\end{eqnarray}
We have introduced the derivative of the chemical potential
with respect to the total number of particles 
($(d\mu/dN)(N=\bar{N})\simeq 2\bar{\mu}/5\bar{N}$
in the Thomas-Fermi limit) in order to recover the characteristic
time scale for the phase collapse of steady state
non mutually interacting condensates.
Our formula reveals the interest of close coupling
constants, such that $g_{aa}+g_{bb}-2g_{ab}\ll g_{aa}$.
In this case
$\chi_d$ is strongly reduced with respect to non mutually interacting 
condensates; $\lambda$ performs small oscillations around the value
$\lambda=1$ so that the integral over $t'$ can be replaced by $t$.
The more general case of close $g$'s not necessarily satisfying the breathe
together condition is analyzed in the next section.

We note that the value of $\chi_s$ as function of $\chi_d$ 
could be expected {\sl a priori} from Eq.(\ref{eq:magique}):
when Eq.(\ref{eq:angle}) is satisfied,
the condensate wavefunctions
form a breathe-together
solution and have therefore a vanishing relative phase
for $N_a=N|c_a|^2,N_b=N|c_b|^2$,
whatever the value of $N$ is.
An important consequence is that there is no extra
damping of the phase coherence due to the fluctuations 
of the total number of particles (see Eq.(\ref{eq:gaus_court})).

\section{Case of close coupling constants}
\label{sec:closeg's}
We consider in this section the case of close coupling constants
which leads to a dramatic reduction of the relative phase decoherence
with respect to the case of non mutually interacting condensates.

The strategy is to solve approximately the Gross-Pitaevskii equations
Eq.(\ref{eq:gpe}) for $\phi_a(N_a,N_b)$ and $\phi_b(N_a,N_b)$ and apply 
the formulas Eq.(\ref{eq:chi_s},\ref{eq:chi_d}) directly. For all equal $g$'s
the initial state is indeed a steady state for the Eq.(\ref{eq:gpe})
and $\chi_s=\chi_d=0$. For close $g$'s we linearize the Gross-Pitaevskii
equations around the initial value in the hydrodynamic point of view. 

\subsection{Linearization in the classical hydrodynamics approximation}
We first rewrite the Gross-Pitaevskii equations Eq.(\ref{eq:gpe}) in terms
of the hydrodynamic variables: 
\begin{eqnarray}
\rho_\varepsilon &\equiv& N_\varepsilon |\phi_\varepsilon(N_a,N_b)|^2 \\
\vec{v_\varepsilon} &\equiv& {\hbar\over m} \; \vec{\mbox{grad}}\; 
\theta_\varepsilon (N_a,N_b)
\end{eqnarray}
that is densities and velocity fields of the two condensates. 
We further assume the Thomas-Fermi limit $\mu \gg \hbar \omega$
and neglect the quantum pressure terms as in \cite{Stringari} in 
the time evolution of the velocity fields: 
\begin{eqnarray}
\partial_t \rho_\varepsilon+\mbox{div}(\rho_\varepsilon\vec{v_\varepsilon}
) & = & 0 
\label{eq:cont_ex}\\
\partial_t\vec{v_\varepsilon} +{1\over 2}\;\vec{\mbox{grad}}\;v_\varepsilon^2
& = & -{1\over m}\;\vec{\mbox{grad}}\;[U(\vec{r})
+\rho_\varepsilon g_{\varepsilon\varepsilon}
+\rho_{\varepsilon'}g_{\varepsilon\varepsilon'}] .
\label{eq:hydro}
\end{eqnarray}
At this point 
we introduce the deviations of the densities and velocity fields
from their initial values: 
\begin{eqnarray}
\rho_\varepsilon (t) & =& \rho_\varepsilon(0) +\delta\rho_\varepsilon(t)\\
\vec{v_\varepsilon} (t) & =& \vec{v_\varepsilon}(0) 
+\delta\vec{v_\varepsilon}(t)
\end{eqnarray}
where the initial values are given by:
\begin{eqnarray}
\rho_\varepsilon(t=0) &=& N_\varepsilon |\phi_0|^2(N) \\
\vec{v_\varepsilon}(t=0) &=& \vec{0}.
\end{eqnarray}
By expanding Eqs.(\ref{eq:cont_ex}),(\ref{eq:hydro}) to first
order in the small quantities 
$\delta\rho_\varepsilon,\delta\vec{v_\varepsilon}$, we obtain:
\begin{eqnarray}
\partial_t \delta\rho_\varepsilon + \mbox{div}[N_\varepsilon |\phi_0|^2
\delta \vec{v_\varepsilon}] & = & 0 
\label{eq:cont}\\
\partial_t \delta \vec{v_\varepsilon} +{1\over m}\;\vec{\mbox{grad}} \;
[\delta\rho_\varepsilon g_{\varepsilon\varepsilon}
+\delta\rho_{\varepsilon'} g_{\varepsilon\varepsilon'}] &=&
-{1\over m}\;\vec{\mbox{grad}}\; [|\phi_0|^2] \times \nonumber\\
&& (N_\varepsilon g_{\varepsilon\varepsilon}+
N_{\varepsilon'}g_{\varepsilon\varepsilon'}-Ng_{aa}).
\label{eq:navier}
\end{eqnarray}
By taking the first time derivative of
Eq.(\ref{eq:cont}) we eliminate the velocity field and we get:
\begin{equation}
\partial_t^2\delta\rho_\varepsilon +\sum_{\varepsilon'}
 M_{\varepsilon\varepsilon'}
S[\delta\rho_{\varepsilon'}]
+\sigma_\varepsilon =0 .
\label{eq:rhoseul}
\end{equation}
The source terms of these inhomogeneous equations are:
\begin{equation}
\sigma_\varepsilon = -{N_\varepsilon\over m}
\mbox{div}[|\phi_0|^2\; \vec{\mbox{grad}}\;|\phi_0|^2]
(N_\varepsilon g_{\varepsilon\varepsilon}+N_{\varepsilon'}
g_{\varepsilon\varepsilon'} - N g_{aa}).
\end{equation}
The homogeneous part of Eq.(\ref{eq:rhoseul}) involves
the $2\times 2$ matrix $M$:
\begin{equation}
M = {1\over N g_{aa}}
\left(\begin{array}{cc} 
N_a g_{aa} & N_a g_{ab} \\
N_b g_{ab} & N_b g_{bb} 
\end{array}
\right)
\end{equation}
and the Stringari operator defined in 
Eq.(\ref{eq:stringari}). 
In order to solve Eq.(\ref{eq:rhoseul}) we introduce the eigenvectors
$\vec{e}_{\pm}$ of the matrix $M$ with corresponding eigenvalues
$g_{\pm}$. Consistently with our previous approximations, we calculate,
to leading order in the differences between the 
coupling constants,
the eigenvalues:
\begin{eqnarray}
g_+ &\simeq& g_{aa} \\
g_- &\simeq& {N_a N_b\over N^2}(g_{aa}+g_{bb}-2g_{ab})
\end{eqnarray}
and the components of $(\delta \rho_a, \delta \rho_b)$ on the 
eigenvectors of $M$:
\begin{eqnarray}
\delta\rho_+ &\simeq& \delta\rho_a + \delta\rho_b \\
\delta\rho_- &\simeq& {N_b\over N}\delta\rho_a - {N_a\over N}\delta\rho_b .
\end{eqnarray}
For those linear combinations we get the decoupled equations:
\begin{equation}
\partial_t^2\delta\rho_{\pm} +{g_\pm\over g_{aa}}
S[\delta\rho_\pm] +\sigma_\pm = 0.
\label{eq:dec}
\end{equation}

To study the dynamics of the system we expand $\rho_{\pm}$ and the
source terms $\sigma_\pm$ on the eigenmodes of the Stringari operator. 
It turns out that the source terms are simply proportional to the
breathing mode $\alpha_{q=5}$ already introduced in Eq.(\ref{eq:mode5}).
The solution of Eq.(\ref{eq:dec}) with the initial conditions
$\delta \rho_{\pm}=\partial_t \delta \rho_{\pm}=0$ is then:
\begin{equation}
\delta\rho_\pm(\vec{r},t) = N|\phi_0(\vec{0})|^2 
{\cal A}_\pm{g_{aa}\over  g_\pm}
[1-\cos\Omega_\pm t] \alpha_{q=5}(r)
\label{eq:res}
\end{equation}
with eigenfrequencies and amplitudes given by:
\begin{eqnarray}
\Omega_\pm &=& \left({5g_\pm\over g_{aa}}\right)^{1/2}\omega\\
{\cal A}_+ &=& {N_a^2 g_{aa} + N_b^2 g_{bb} +2 N_a N_b g_{ab}\over N^2 g_{aa}}
-1\\
{\cal A}_- &=&  {N_a N_b\over N^2}
\left[{N_a g_{aa}+N_b g_{ab}-N_b g_{bb}-N_a g_{ab}\over
N g_{aa}}\right].
\end{eqnarray}
We note that when the numbers of atoms $N_{a,b}$
satisfy the breathe-together condition Eq.(\ref{eq:angle})
the amplitude ${\cal A}_-$ vanishes as expected, since $\delta\rho_-
\equiv 0$ in this case.

\subsection{Validity of the linear approximation}
In order for our linearized treatment to be valid the deviations
$\delta \rho_{\pm}$ should remain small as compared to the initial densities.
A first necessary condition to be satisfied is that the eigenfrequencies
$\Omega_\pm$ should be real. This imposes the positivity of the matrix $M$,
ensured by the positivity of its determinant:
\begin{equation}
g_{ab}^2 \leq g_{aa} g_{bb} .
\label{eq:stab}
\end{equation}
This condition is known in the case of homogeneous mixtures of condensates 
as a stability condition against demixing \cite{homogene}.
To the leading order in the difference between the coupling constants,
the condition Eq.(\ref{eq:stab}) is equivalent to $g_{aa}+g_{bb}-2g_{ab}>0$.

We note at this
point that the amplitude ${\cal A}_-/g_-$ in the
expression for $\delta \rho_-$ is a ratio
of two small numbers. 
When this ratio is  large the system can evolve far from its initial
state even in the stable case $g_->0$:
numerical solutions of the Gross-Pitaevskii equations confirm
this expectation, showing  the formation
of a crater at the center of one of the condensates.
We therefore have to impose a second condition:
\begin{equation}
|{\cal A}_\pm{g_{aa}\over  g_{\pm}}| \ll 1 .
\label{eq:seconde_cond}
\end{equation}

Finally the present treatment is based on the classical hydrodynamic 
approximation;
by including the quantum pressure terms 
in the hydrodynamic
equation for the velocity field
one can show that
this imposes on the eigenfrequencies $\Omega_-$:
\begin{equation}
\frac{\hbar \omega^2}{\mu} \ll \Omega_{-}
\end{equation}
(see also Appendix \ref{app:TF}).
This condition can be violated even in the Thomas-Fermi limit,
when the $g_-$ eigenvalue almost vanishes. In this case one has to
include the quantum pressure terms;
the decoupling property of $\delta \rho_\pm$
is unaffected; for the evolution of $\delta \rho_-$ similar results
as in Eq.(\ref{eq:res}) are obtained; we find e.g.\ $\Omega_-\simeq
63\hbar\omega^2/8\mu$.

\subsection{Phase dynamics}
We assume that all the conditions for the validity of the linearized
treatment are satisfied so that we can proceed
to the analysis of the relative phase dynamics.
To this aim we write the equation of evolution
for the phases $\theta_\varepsilon$ of the condensate wavefunctions 
$\phi_\varepsilon$ in the classical hydrodynamic approximation:
\begin{equation}
\partial_t \theta_\varepsilon +{\hbar\over 2m}
\left(\;\vec{\mbox{grad}}\;\theta_\varepsilon\right)^2= - [ U + g_{\varepsilon \varepsilon}
	\rho_\varepsilon + g_{\varepsilon \varepsilon'}
        \rho_{\varepsilon' }]/\hbar .
\label{eq:theta_hydro}
\end{equation}
The equations for the velocity fields previously given are simply the
gradient of Eq.(\ref{eq:theta_hydro}).
By linearizing Eq.(\ref{eq:theta_hydro}) around the initial state
$\theta_\varepsilon =0$ we obtain for the relative phase:
\begin{eqnarray}
\hbar 
\partial_t(\theta_a- \theta_b) &\simeq& 
- |\phi_0|^2 ( N_a g_{aa} + N_b g_{ab}
  - N_b g_{bb} - N_a g_{ab})\\
& +& (g_{ab} - g_{aa}) \delta \rho_a +
  (g_{bb} - g_{ab}) \delta \rho_b . 
\end{eqnarray}
The right hand side of this equation is a sum of terms constant in time
and of oscillatory functions of time.
The function $\theta_a-\theta_b$
then has two components: an oscillating component and a component diverging
linearly with time which will dominate for long times. 
By using the result Eq.(\ref{eq:res}) and the Thomas-Fermi approximation
for $|\phi(0)|^2$ Eq.(\ref{eq:phi_TF})
we can calculate the time diverging component and
we obtain to leading order in the $g$'s difference:
\begin{equation}
\theta_a - \theta_b \sim - \frac{2 \mu}{5 N g_{aa}} [N_a g_{aa} - N_b g_{bb}
	+ (N_b -N_a) g_{ab} ] \; t/\hbar .
\label{eq:evol_theta_hydro}
\end{equation}
We now use Eq.(\ref{eq:chi_d}) and Eq.(\ref{eq:magique}) to obtain: 
\begin{eqnarray}
\chi_d &\sim & -{1\over 2} \left({d\mu\over dN}\right)_{N=\bar{N}}
 {g_{aa}+g_{bb}
-2 g_{ab}\over g_{aa}}\;t/\hbar \\
\chi_s+(|c_a|^2-|c_b|^2)\chi_d & \sim& - \frac{2 }{5 g_{aa}} 
\left(\frac{d \mu}{dN}\right)_{N=\bar{N}} \times\nonumber \\
&& (|c_a|^2 g_{aa} + |c_b|^2 g_{ab} -|c_b|^2 g_{bb} - |c_a|^2 g_{ab}) \;t/\hbar 
\end{eqnarray}
where we introduced the derivative of the chemical potential with respect
to the total number of particle $(d\mu/dN)(N=\bar{N}) 
\simeq (2/5) \bar{\mu}/\bar{N}$ 
in the Thomas-Fermi limit.
As we already found in the particular case of the breathe-together solution
the constants $\chi_d$ and $\chi_s$
governing the relative phase collapse are
highly reduced for close $g$'s with respect to the case
of non mutually interacting condensates.

\subsection{Physical interpretation of the results}
We now show that all the previous results of this section can be interpreted
in terms of small oscillations of the condensates around the
steady state.

Let us introduce the steady state densities 
$\rho_{\varepsilon}^{\mbox{\scriptsize st}}$ for the condensates
with $N_a$ particles in $a$ and $N_b$ particles in $b$. 
As we are in the case of quasi complete spatial overlap between
the two condensates we can use the Thomas-Fermi approximation
to determine these densities:
\begin{eqnarray}
\mu_a -U &=& \rho_a^{\mbox{\scriptsize st}} g_{aa} + 
\rho_b^{\mbox{\scriptsize st}} g_{ab} \\
\mu_b -U &=& 
\rho_a^{\mbox{\scriptsize st}} g_{ab}  +
\rho_b^{\mbox{\scriptsize st}} g_{bb} 
\end{eqnarray}
where $\mu_\varepsilon$ are the chemical potentials in steady
state.
We rewrite these equations in terms of the deviations
$\delta\rho_{\varepsilon}^{\mbox{\scriptsize st}}$ of
the steady-state densities from the initial state
densities $N_\varepsilon|\phi_0|^2$ and in terms of
the deviations $\delta\mu_\varepsilon$ of the chemical potentials
from $\mu$ defined in Eq.(\ref{eq:phi0}):
\begin{eqnarray}
\label{eq:syst_sta1}
\delta\mu_a &= & (N_a g_{aa}+N_b g_{ab}-Ng_{aa}) |\phi_0|^2 
+\delta\rho_a^{\mbox{\scriptsize st}} g_{aa}
+\delta\rho_b^{\mbox{\scriptsize st}} g_{ab} \\
\delta\mu_b &= & (N_b g_{bb}+N_a g_{ab}-Ng_{aa}) |\phi_0|^2 
+\delta\rho_a^{\mbox{\scriptsize st}} g_{ab}
+\delta\rho_b^{\mbox{\scriptsize st}} g_{bb}.
\label{eq:syst_sta2}
\end{eqnarray}
Using the fact that the spatial integral of $\delta\rho_\varepsilon$
vanishes, we get from integration of Eq.(\ref{eq:syst_sta1},\ref{eq:syst_sta2})
over the volume of $|\phi_0|^2$ the approximate relations:
\begin{eqnarray}
\delta\mu_a &=& {2\mu\over 5Ng_{aa}}(N_a g_{aa} + N_b g_{ab} -N g_{aa}) \\
\delta\mu_b &=& {2\mu\over 5Ng_{aa}}(N_b g_{bb} + N_a g_{ab} -N g_{aa}).
\end{eqnarray}
We can therefore check that the relative phase of the condensates in steady 
state, given by $\theta_a^{\mbox{\scriptsize st}}-
\theta_b^{\mbox{\scriptsize st}}= -i(\delta\mu_a-\delta\mu_b)t/\hbar$,
evolves as in Eq.(\ref{eq:evol_theta_hydro}). The phase decoherence
properties of the evolving mixture are then essentially the same 
as in steady state.

Moreover we now show that the average $\langle\delta\rho_\varepsilon
\rangle$ 
of $\delta\rho_\varepsilon$ over the oscillations at frequencies
$\Omega_\pm$  coincide with $\delta\rho_\varepsilon^{\mbox{\scriptsize st}}$.
First, by averaging Eq.(\ref{eq:cont}) over time we find
that the velocity fields have a vanishing time average \cite{details}.
Second, we average Eq.(\ref{eq:navier}) over time; we find 
equations for the spatial gradient of $\langle\delta\rho_\varepsilon
\rangle$, which
coincide with the spatial gradient of Eq.(\ref{eq:syst_sta1},
\ref{eq:syst_sta2}), so that
$\langle\delta\rho_\varepsilon\rangle
=\delta\rho_\varepsilon^{\mbox{\scriptsize st}}$ \cite{proviso2}.

\section{Discussion of the JILA case}
\label{sec:JILA}

In the JILA experiment the values of the three coupling constants
between the atoms are known with good precision; they are in
the ratio  \cite{JILA_demix}:
\begin{equation}
g_{aa} : g_{ab} : g_{bb} = 1.03\ :\ 1\ :\ 0.97 .
\end{equation}
No breathe-together solution exists in this case, as $g_{ab}$ lies
within $g_{aa}$ and $g_{bb}$. Experimentally half of the particles
are in the state $a$ so that $|c_a|^2=|c_b|^2=1/2$, and the mean
total number of particles is $\bar{N}=5\times 10^5$.
Although the coupling constants are close, the linearized treatment
presented in section \ref{sec:closeg's} does not apply either,
because condition Eq.(\ref{eq:seconde_cond}) is violated.
It is actually found experimentally that the two condensates
evolve far from the initial state, with formation of a crater
in the $a$ condensate while the $b$ condensate becomes more confined
at the center of the trap;
eventually the condensates separate in some random direction
\cite{JILA_demix}.

To avoid the crater formation and trigger the spatial separation of the two
condensates in a reproducible direction  a small spatial shift
is applied to the trapping potential of one of the two states.
The two condensates separate, with a relative motion exhibiting
strongly damped oscillations \cite{JILA_demix}. The system then reaches
a steady state that still exhibits phase coherence, up to times
on the order of $150$ ms after the phase state preparation
\cite{JILA_phase}.

\subsection{Time dependent calculations}
We have already studied in \cite{Nous} the damping of the relative motion
between the condensates, by numerical integration of the coupled
Gross-Pitaevskii equations Eq.(\ref{eq:gpe}). The agreement with
the experimental results of \cite{JILA_demix} is qualitatively
good, although the damping in the theory is weaker and incomplete,
small oscillations of the condensate wavefunctions remaining undamped 
even at long times.

We have applied the formalism of section \ref{sec:gen-method} by
numerically integrating the Gross-Pitaevskii equations for the
parameters of the JILA experiment. The coefficients $\chi_s,\chi_d$,
now complicated functions of time and space,
are obtained by evolving wavefunctions with slightly different
numbers of atoms in $a$ and $b$. In order to facilitate the
comparison with the experiments, in which the $x$-integrated atomic
density $\bar{\rho}_a(y,z)$ in the internal state $a$
is measured after the $\pi/2$ pulse applied at time
$\tau$, we calculated the following
contrast:
\begin{equation}
C_{\mbox{\scriptsize JILA}}(y,z) =
{ \mbox{max}_\delta \bar{\rho}_a -\mbox{min}_\delta\bar{\rho}_a
\over
\mbox{max}_\delta\bar{\rho}_a
+\mbox{min}_\delta\bar{\rho}_a}
=
{2|\int dx\;\langle\hat{\psi}_a^\dagger(\tau^-)\hat{\psi}_b(\tau^-)\rangle
^{\mbox{\scriptsize Gauss}}|
\over \sum_{\varepsilon=a,b}
\int dx\;  \bar{N_\varepsilon}|\bar{\phi_\varepsilon}|^2
(\tau^-)}
\end{equation}
where the interference term Eq.(\ref{eq:inter_N}) is averaged over
a Gaussian distribution of the total number of particles
with a standard deviation $\Delta N$.
A direct comparison with the
experiment would require the inclusion of the 22 ms ballistic
expansion, not included in the present numerical calculations.

Our numerical result for $C_{\mbox{\scriptsize JILA}}$
at the center of the trap for the species $a$, $y=z=0$, is presented in
Fig.\ref{fig:JILAa}, for Gaussian fluctuations
in the total number of particles $\Delta N/\bar{N}
=8 \%$ corresponding to the JILA experiment
\cite{private}, together with the pure Gross-Pitaevskii
prediction $C_{\mbox{\scriptsize GPE}}$
obtained by setting all the $\chi$'s to 0. 
The Gross-Pitaevskii prediction oscillates around
$\langle C_{\mbox{\scriptsize GPE}}\rangle =0.63$.
On the contrary the result of the more complete calculation 
including fluctuations in the relative and total number
of particles exhibits a damping of the contrast, that we have
fitted by convenience with the formula $C_{\mbox{\scriptsize JILA}}
=C_0 e^{-\gamma t}$; we obtain $C_0\simeq 
\langle C_{\mbox{\scriptsize GPE}}\rangle$ and $\gamma^{-1}=0.42$s.

Note the oscillatory aspect of the curves in Fig.\ref{fig:JILAa}.
More understanding of the structure of the condensate wavefunctions
given by Eq.(\ref{eq:gpe}) is required as this point: as detailed in
\cite{Nous} $\bar{\phi_\varepsilon}$ is a sum of a smooth part, 
performing oscillations with frequencies expected to be
close to eigenfrequencies of the
steady state condensates \cite{verif}, and of a noisy quasi-stochastic part.
The slow oscillatory structure evident on 
$C_{\mbox{\scriptsize GPE}}$ comes from this smooth oscillating part
of the wavefunctions.

We have also considered the ideal case of a well defined
total number of particles. The numerical prediction 
for the contrast $C_{\mbox{\scriptsize JILA}}$ in this case
corresponds to a very long lived phase coherence: after a time of 
1 second, the contrast is still very close to 
the pure Gross-Pitaevskii prediction.

\begin{figure}
\caption{For the parameters of the JILA experiment (not
including the 22 ms ballistic
expansion),  phase contrasts 
$C_{\protect\mbox{\protect\scriptsize JILA}}$
(lower curve) and 
$C_{\protect\mbox{\protect\scriptsize GPE}}$ 
(upper curve) defined in
the text, at $y=z=0$, as function of time in seconds, for the
evolving binary mixture, with $\Delta N=0.08\bar{N}$.
\label{fig:JILAa}}
\end{figure}

\subsection{Steady state calculations and effect of particle losses}
As the wavefunctions at long times
perform mainly oscillations around the
steady state we have also tried a much simpler steady
state calculation (see subsection \ref{subsec:sta}).
During the collapse time the contrast $C_{\mbox{\scriptsize JILA}}$
is a Gaussian in time Eq.(\ref{eq:gaus_court}), with an initial
value $0.958$ and
with a half-width $t_c$ at the relative height $e^{-1/2}$.
We plot in Fig.\ref{fig:JILAb} the variation of $t_c$ as function
of the standard deviation $\Delta N$.
As we find $\chi_s/t=-7.7\times 10^{-5} $ s${}^{-1} $ and $\chi_d/t= 
-4.5\times 10^{-4}$ s${}^{-1} $,
one has $|\chi_s|\simeq |\chi_d|/6$, so that
relatively high values of $\Delta N$ are required to observe a significant
effect of the fluctuations of the total number of particles on
phase decoherence.
For $\Delta N =0.08 \bar{N}$
the phase decoherence time is $t_c=0.32$ second, close to the
result of the time-dependent calculation of Fig.\ref{fig:JILAa}.
Note that for such a high value of $\Delta N/\bar{N}$ the
decay of the phase contrast in Eq.(\ref{eq:gaus_court}) is
essentially due to the first exponential factor accounting for
the smearing of the phase by fluctuations of the total number
of particles, the spreading of the phase for a fixed number of 
particles being very small ($\bar{N}\chi_d^2(t_c)/2\simeq 0.005$).

We now briefly consider the issue of losses of particles.
An intrinsic source of losses in the JILA experiment are
the inelastic collisions between $a$ atoms and $b$ atoms,
resulting in the simultaneous loss of two particles.
We estimate the mean number $\langle\delta N\rangle$ of lost particles 
from the rate constant $K_2$
for binary inelastic collisions between the states $|F=1,m=-1\rangle$
and $|F=2,m=2\rangle$ \cite{overlap} and from a numerical calculation
of the overlap integral $\int d^3\vec{r}\,|\bar{\phi_a}|^2|\bar{\phi_b}|^2$.
For the JILA parameters we find $\langle\delta N\rangle/\bar{N}=0.04$ at time
$t_c=0.32$ second. One could then naively expect the effect of
losses on phase coherence to be
comparable with the effect of fluctuations of $N$.

To test this naive expectation we use the following simple model,
inspired by the two-mode model developed in \cite{resur}, and focusing
on the effect of the losses on the drift velocity $v(N)$ of the relative
phase of the two condensates given in Eq.(\ref{eq:vdrift}).
Imagine that the system has initially $\bar{N}$ condensate
atoms and that $k$ binary inelastic collisions have taken place
at times $t_1 < \ldots <t_k$ between time $0$ and time $t$. The shift
of the relative phase during $t$ is then given by:
\begin{equation}
\Theta = \int_0^t d\tau\; v(N(\tau))  = v(\bar{N}) t_1 + v(\bar{N}-2) (t_2-t_1) +\ldots +
v(\bar{N}-2k)(t-t_k).
\end{equation}
As we do in \cite{resur} we assume a constant mean number of
collisions $\lambda$ per unit of time and
we average the phase factor $e^{i\Theta}$ multiplying the
interference term $\langle\hat{\psi_b}^\dagger\hat{\psi_a}\rangle$
over the probability distribution of the times $t_1,\ldots,
t_k$ and of the number of loss events $k$, 
\begin{equation}
P_t(t_1,\ldots,t_k;k) = \lambda^k e^{-\lambda t}
\end{equation}
to obtain:
\begin{eqnarray}
|\langle e^{i\Theta}\rangle| &=& \exp
\left\{-\langle k\rangle[1-\sin(2\chi_s)/
(2\chi_s)]\right\} \\
&\simeq &  \exp
\left[-{2\over 3}\langle k\rangle\chi_s^2\right] \ \ \ \ \mbox{for}
\ |\chi_s| \ll 1
\label{eq:perte_court}
\end{eqnarray}
where $2\langle k\rangle=2\lambda t=\langle \delta N\rangle$ 
is the mean number of lost particles during $t$.
At time $t=t_c=0.32$ second the corresponding modulus of
the averaged phase factor is on the order of $[1-4\times 10^{-6}]$,
very close to one: particle losses have a negligible effect
on the phase coherence at the considered time $t_c$, even
if $\langle\delta N\rangle$ and $\Delta N$ have the same order of magnitude.

Actually an inspection of the $\chi_s$ dependent factor
in Eq.(\ref{eq:gaus_court})
and of Eq.(\ref{eq:perte_court}) reveals that these equations
have the same structure; replacing in Eq.(\ref{eq:gaus_court})
the variance $\Delta N^2$ of the total number of particles by
the variance $\Delta k^2$ of the number of loss events 
($\Delta k^2=\langle k\rangle$ as $k$ obeys a Poisson law)
one recovers Eq.(\ref{eq:perte_court}) up to a numerical factor
inside the exponential. For equally large values of $\Delta N$ and $\langle
k\rangle$ the effect of losses on phase coherence is less important
than that of fluctuations of $N$ because $\Delta k^2 = \langle k\rangle
\ll \Delta N^2$.

We have also investigated another source of
losses, the collisions of condensate atoms
with the background gas of the cell.
Assuming a lifetime of the particles in the cell of 250 seconds as in
\cite{decay} we find as well that this loss mechanism has a negligible
effect on the phase coherence for a time $t_c=0.32$ second.

\begin{figure}
\caption{For the parameters of the JILA experiment (except the 22 ms ballistic
expansion),
collapse time $t_c$ for
$C_{\protect\mbox{\protect\scriptsize JILA}}$ at $y=z=0$ as a
function of $\Delta N/\bar{N}$
for zero temperature steady state condensates in the shifted
traps.
\label{fig:JILAb}}
\end{figure}
\section{Conclusion and perspectives}

We have extended previous treatments of the phase dynamics of Bose-Einstein
condensates at zero temperature
to the case of mutually interacting and dynamically evolving
binary mixtures of condensates, for a measurement scheme
of the phase coherence inspired by the JILA experiment.

We have first applied this extended formalism to the interesting
breathe-together solution of the Gross-Pitaevskii equations,
in which the two condensates oscillate in phase, remaining always
exactly spatially superimposed. 
The analytical results for the phase show that a dramatic increase
of the phase coherence time can be obtained for close coupling
constants $g_{aa},g_{ab},g_{bb}$ describing the elastic
interactions between $a$ atoms and $b$ atoms.

We have also treated analytically the case of close $g$'s,
in the absence of demixing instability. Basically the phase
collapse is identical to the steady state case for the two 
mutually interacting condensates.

Finally, we have investigated numerically the more difficult case 
of JILA. We find a collapse time of the phase on the order of
0.4 second, both by a dynamical and a steady state calculation,
in the case of Gaussian fluctuations of the total number of particles,
corresponding to $\Delta N/\bar{N} = 8\%$.
This result for the collapse time is significantly larger
than the experimental results (no phase
coherence measured after 150 ms). We have also estimated in a simple way
the effect of collisional losses on phase coherence in the JILA
experiment.

A possible extension of this work could include the effect of
the presence of a thermal component in the experiment.

\section*{Acknowledgements}

Part of this work (the breathe-together solution)
would have not been possible without
the contribution of G.\ Shlyapnikov, J.\ Dalibard and 
P.\ Fedichev.  We thank A.\ Leggett, Y.\ Kagan for very useful discussions
on the role of fluctuations in the total number of particles.
We thank Ralph Dum for help in the numerical calculations.
A.\ S.\ acknowledges financial support from the European Community
(TMR individual research grant).

\appendix
\section{Phase correction to the Gross-Pitaevskii prediction}
\label{app:A}

We consider the evolution of the Fock state $|N_a:\phi_a(0),N_b:\phi_b(0)
\rangle$ (with $N_{a,b}$ particles in the internal state $a,b$).
The model Hamiltonian we consider contains the one-body Hamiltonians
${\cal H}_\varepsilon$ and elastic interactions terms:
\begin{eqnarray}
H &=& \int d^3\vec{r}\ \sum_{\varepsilon=a,b}\hat{\psi_\varepsilon}^\dagger
{\cal H}_\varepsilon\hat{\psi_\varepsilon} + \nonumber \\
&& {1\over 2} g_{aa}\hat{\psi_a}^\dagger\hat{\psi_a}^\dagger
\hat{\psi_a}\hat{\psi_a}
+ {1\over 2} g_{bb}\hat{\psi_b}^\dagger\hat{\psi_b}^\dagger
\hat{\psi_b}\hat{\psi_b}
+ g_{ab}\hat{\psi_b}^\dagger\hat{\psi_a}^\dagger\hat{\psi_a}\hat{\psi_b}
\end{eqnarray}
where $\hat{\psi_\varepsilon}$ is the atomic field operator in the internal
state $\varepsilon$.

We use the Hartree-Fock type ansatz for the $N$-body state vector:
\begin{equation}
|\Psi\rangle = e^{-iA(t)/\hbar} |N_a:\phi_a(t),N_b:\phi_b(t)\rangle.
\end{equation}
A variational formulation of the Hamiltonian equation 
\begin{equation}
i\hbar {d\over dt}|\Psi\rangle = H |\Psi\rangle
\label{eq:schrod}
\end{equation}
leads to the Gross-Pitaevskii equations for $\phi_\varepsilon(t)$,
given in Eq.(\ref{eq:gpe}), up to the undetermined
phase factor $A$ corresponding formally to a time dependent Lagrange
multiplier ensuring the conservation of the norm of $|\Psi\rangle$. 
To determine this phase factor $A$, we multiply Eq.(\ref{eq:schrod})
on the left by the bra $\langle\Psi|$; we obtain:
\begin{equation}
\dot{A}+i\hbar \langle N_a:\phi_a(t),N_b:\phi_b(t)|{d\over dt}
|N_a:\phi_a(t),N_b:\phi_b(t)\rangle = \langle \Psi | H|\Psi\rangle.
\end{equation}
The scalar products are calculated in second quantized formalism,
e.g.\ we find:
\begin{equation}
\langle N_a:\phi_a(t),N_b:\phi_b(t)|{d\over dt}
|N_a:\phi_a(t),N_b:\phi_b(t)\rangle =
\sum_\varepsilon N_\varepsilon 
\langle\phi_\varepsilon|{d\over dt}|\phi_\varepsilon\rangle.
\end{equation}
We finally arrive at Eq.(\ref{eq:apoint}).

\section{Derivation of the interference term}
\label{app:interf}

When the $N$-body state vector is initial a phase
state Eq.(\ref{eq:init}) and if one assumes that the
Fock states evolve according to Eq.(\ref{eq:evol_Fock})
one gets the following expression for the interference term
between the two condensates:
\begin{eqnarray}
\langle \hat{\psi}_b^\dagger \hat{\psi_a}\rangle _{N}
&=& c_a c_b^* \sum_{N_a=1}^{N} {N!\over (N_a-1)!N_b!}
|c_a|^{2(N_a-1)}|c_b|^{2N_b}
\phi_a(N_a,N_b)\phi_b^*(N_a-1,N_b+1)\times \nonumber\\
&&e^{i[A(N_a-1,N_b+1)-A(N_a,N_b)]/\hbar} \times \nonumber \\
&&[\langle \phi_a(N_a-1,N_b+1)|\phi_a(N_a,N_b)\rangle]^{N_a-1}
[\langle \phi_b(N_a-1,N_b+1)|\phi_b(N_a,N_b)\rangle]^{N_b}
\end{eqnarray}
where $N_b=N-N_a$.
In the large $N$ limit, we expand to first order the effect 
of shifts of $N_{\varepsilon}$ by unity in the last two lines
of the previous equation:
\begin{eqnarray}
\phi_a(N_a-1,N_b+1) &\simeq& \phi_a(N_a-1,N_b) +\partial_{N_b}
\phi_a(N_a-1,N_b) \\
\phi_a(N_a,N_b) &\simeq& \phi_a(N_a-1,N_b) +\partial_{N_a}
\phi_a(N_a-1,N_b) \\
A(N_a-1,N_b+1) &\simeq & A(N_a-1,N_b) +\partial_{N_b} A(N_a-1,N_b) \\
A(N_a,N_b) &\simeq & A(N_a-1,N_b) +\partial_{N_a} A(N_a-1,N_b).
\end{eqnarray}
We then get:
\begin{eqnarray}
\langle \hat{\psi}_b^\dagger \hat{\psi_a}\rangle _{N}
&=& c_a c_b^* \sum_{N_a=1}^{N} {N!\over (N_a-1)!N_b!}
|c_a|^{2(N_a-1)}|c_b|^{2N_b}
\phi_a(N_a,N_b)\phi_b^*(N_a-1,N_b+1)\times \nonumber\\
&& e^{i\Theta(N_a-1,N_b)}
\end{eqnarray}
where we have introduced the real quantity:
\begin{equation}
\Theta(N_a,N_b) = {1\over\hbar}
(\partial_{N_b}-\partial_{N_a})A(N_a,N_b)
+i\left[\sum_\varepsilon N_\varepsilon
\langle\phi_\varepsilon(N_a,N_b)|(\partial_{N_b}-\partial_{N_a})
|\phi_\varepsilon(N_a,N_b)\rangle\right].
\end{equation}
We calculate the time derivative of $\Theta(N_a,N_b)$ using the 
Gross-Pitaevskii equations Eq.(\ref{eq:gpe}). After lengthy 
calculations we find
\begin{equation}
\dot{\Theta}(N_a,N_b) = 0.
\end{equation}
In the gedanken experiment considered in this paper, the
initial wavefunctions $\phi_\varepsilon(t=0)$ depend
only on $N_a+N_b$ so that they have a vanishing derivative
$\partial_{N_b}-\partial_{N_a}$, and we take initially $A$=0;
this leads to $\Theta\equiv 0$.
The same conclusion holds if the initial wavefunctions are real.

\section{Approximate evolution in the Thomas-Fermi limit}
\label{app:TF}

After the gauge and scale transforms Eq.(\ref{eq:gau1},\ref{eq:gau2})
the equations of evolution for $\tilde{\bar{\phi}}$ and
$\tilde{\delta\varphi_d}$ read:
\begin{eqnarray}
i\hbar\partial_t\tilde{\bar{\phi}} &=&-{\hbar^2\over 2m\lambda^2}
\Delta\tilde{\bar{\phi}} + {g\over g_{aa}\lambda^3}\left[U(\vec{r})
+\bar{N}g_{aa} |\tilde{\bar{\phi}}|^2-\bar{\mu}\right]\tilde{\bar{\phi}}
\label{eq:evol1}\\
i\hbar\partial_t \tilde{\delta\varphi_d} &=& 
-{\hbar^2\over 2m\lambda^2}\Delta\tilde{\delta\varphi_d} 
+{g\over g_{aa}\lambda^3}
\left[U(\vec{r})+\bar{N}g_{aa} |\tilde{\bar{\phi}}|^2-\bar{\mu}\right] 
\tilde{\delta\varphi_d}\nonumber\\
&+&
{1\over\lambda^3} N_b(g_{bb}-g_{ab})
(|\tilde{\bar{\phi}}|^2\tilde{\delta\varphi_d} + \tilde{\bar{\phi}}^2
\tilde{\delta\varphi_d^*}).\label{eq:evol2}
\end{eqnarray}

In the Thomas-Fermi limit the terms involving the Laplacian are 
small; if we neglect them we get for the time derivatives
of the $\alpha$ and $\beta$
variables defined in Eq.(\ref{eq:def_alpha},\ref{eq:def_beta}):
\begin{eqnarray}
i\hbar \partial_t \alpha &=& 0 
\label{eq:pas_bon}\\
i\hbar \partial_t \beta &=& {1\over \lambda^3} N_a(g_{aa}-g_{ab})
\alpha.
\label{eq:suffit}
\end{eqnarray}
The variable $\alpha$ has actually been defined in a way to obtain
zero on the right hand side of Eq.(\ref{eq:pas_bon}).

The first equation Eq.(\ref{eq:pas_bon})
is not an acceptable approximation for the evolution
of $\alpha$, we therefore include in $\partial_t \alpha$
the contribution of the Laplacian terms:
\begin{equation}
i\hbar \partial_t \alpha = -{\hbar^2\over 2m\lambda^2} 
\;\mbox{div}\;\left\{
\alpha\left[{\;\vec{\mbox{grad}}\;\tilde{\bar{\phi}}\over\tilde{\bar{\phi}}}
-{\;\vec{\mbox{grad}}\;\tilde{\bar{\phi}}^*\over\tilde{\bar{\phi}}^*}\right]
+2 |\tilde{\bar{\phi}}|^2\;\vec{\mbox{grad}}\;\beta\right\}.
\end{equation}
Furthermore, along the lines of reference \cite{Yvan}, one
can show that $\tilde{\bar{\phi}}$ has a negligible time evolution
in the Thomas-Fermi limit; we can then replace $\tilde{\bar{\phi}}$
by its initial value $\bar{\phi_0}$ and we recover the first line
of Eq.(\ref{eq:evol_ab}).

The second equation Eq.(\ref{eq:suffit}) is an acceptable approximation
for the evolution of $\beta$ if the neglected terms, all involving 
spatial derivatives of $\alpha,\beta$ or $\tilde{\bar{\phi}}$, are small
as compared to the right hand side of Eq.(\ref{eq:suffit}),
as they are expected to be in the Thomas-Fermi limit. Neglecting these terms,
we recover the second line of Eq.(\ref{eq:evol_ab}).

In order to estimate the order of magnitude of the neglected
terms in the time derivative of $\beta$, we calculate the exact
derivative:
\begin{eqnarray}
i\hbar \partial_t \beta &=& {1\over \lambda^3} N_a(g_{aa}-g_{ab}) \nonumber\\
&& -{\hbar^2\over 2m\lambda^2} \;\vec{\mbox{grad}}\;\beta
\cdot\left[{\;\vec{\mbox{grad}}\;\tilde{\bar{\phi}}\over\tilde{\bar{\phi}}}
-{\;\vec{\mbox{grad}}\;\tilde{\bar{\phi}}^*
\over\tilde{\bar{\phi}}^*}\right] \nonumber \\
&& -{\hbar^2\over 2m\lambda^2|\tilde{\bar{\phi}}|^2}
\left\{\Delta\alpha-{1\over 2}\;\vec{\mbox{grad}}\;\alpha\cdot
\left[{\;\vec{\mbox{grad}}\;\tilde{\bar{\phi}}\over\tilde{\bar{\phi}}}
+{\;\vec{\mbox{grad}}\;\tilde{\bar{\phi}}^*\over\tilde{\bar{\phi}}^*}\right]
-{1\over 2}\alpha\left[{\Delta\tilde{\bar{\phi}}\over\tilde{\bar{\phi}}}+
{\Delta\tilde{\bar{\phi}}^*\over\tilde{\bar{\phi}}^*}\right]\right\}.
\end{eqnarray}
We replace $\tilde{\bar{\phi}}$ by $\bar{\phi_0}$. We consider an eigenmode
with frequency $\Omega_q$; from Eq.(\ref{eq:eigen}) we
estimate $\Delta\alpha/\alpha \sim q/R_0^2$.
Assuming $\lambda$ on the order of $1$ we get the condition
\begin{equation}
\Omega_q \gg q{\hbar\omega^2\over\mu}
\end{equation}
which we can rewrite as
\begin{equation}
1\leq\sqrt{q}\ll {\Omega_{q=1}\over \hbar\omega^2/\mu}.
\label{eq:borne_q}
\end{equation}


\begin{thebibliography}{99}
\bibitem{demons}
Consider the case $\Omega_q = i\gamma$, where $\gamma>0$;
then $A_q,B_q$ are real positive for all time and one has:
\begin{equation}
{d\over dt}(A_qB_q) ={\gamma\over\lambda^2}B_q^2+{\gamma\over\lambda^3}A_q^2
\geq {\gamma\over\lambda_{\mbox{\scriptsize max}}^3}(A_q^2+B_q^2)
\geq {2\gamma\over\lambda_{\mbox{\scriptsize max}}^3} A_qB_b
\end{equation}
where $\lambda_{\mbox{\scriptsize max}}=\mbox{max}_t \lambda(t)\geq 1$.
\bibitem{Alice_est_contre} The calculation was done 
for $g_{ab}/g_{aa}=0.4596$,
$g_{bb}/g_{aa}=0.67$ (leading to $\sigma=2.6\times 10^{-2}\omega$)
with a chemical potential $\bar{\mu}\simeq 72.5\hbar\omega$
and a deviation $\delta N_a = -\delta N_b=-0.1 \bar{N_a}$.
\bibitem{proviso} We do not get the linear divergence with
time of the quadrupolar mode $q=5$ predicted in \cite{Ralph}
as the equations for $\delta\varphi_d$ differ from the ones
obtained for a single condensate, leading to a frequency of the breathing
eigenmode $\Omega_5$ very different from the one of a single
condensate.
The divergence of the quadrupolar mode occurs here
for $\delta\varphi_s$.
\bibitem{details} More precisely we arrive at $\mbox{div}[|\phi_0|^2
\langle\delta\vec{v_\varepsilon}\rangle]=0$. The velocity fields
are of the form $f_\varepsilon(r)\vec{r}$ because the condensate wavefunctions
are isotropic. Using Gauss formula we find that $f_\varepsilon\equiv 0$.
\bibitem{proviso2} More precisely  $\langle\delta\rho_\varepsilon
\rangle$ and $\delta\rho_\varepsilon^{\mbox{\scriptsize st}}$
differ by a constant; on the other hand this constant vanishes as both the
density deviations have a vanishing integral over the volume
of the condensate, so we conclude $\langle\delta\rho_\varepsilon\rangle
=\delta\rho_\varepsilon^{\mbox{\scriptsize st}}$.
\bibitem{verif} We have checked numerically this statement 
for the JILA case by looking at the time dependence of
the relative distance between
the two condensates; at long times this distance oscillates at a
frequency close to an eigenfrequency of the steady state
condensates.
\end{thebibliography}
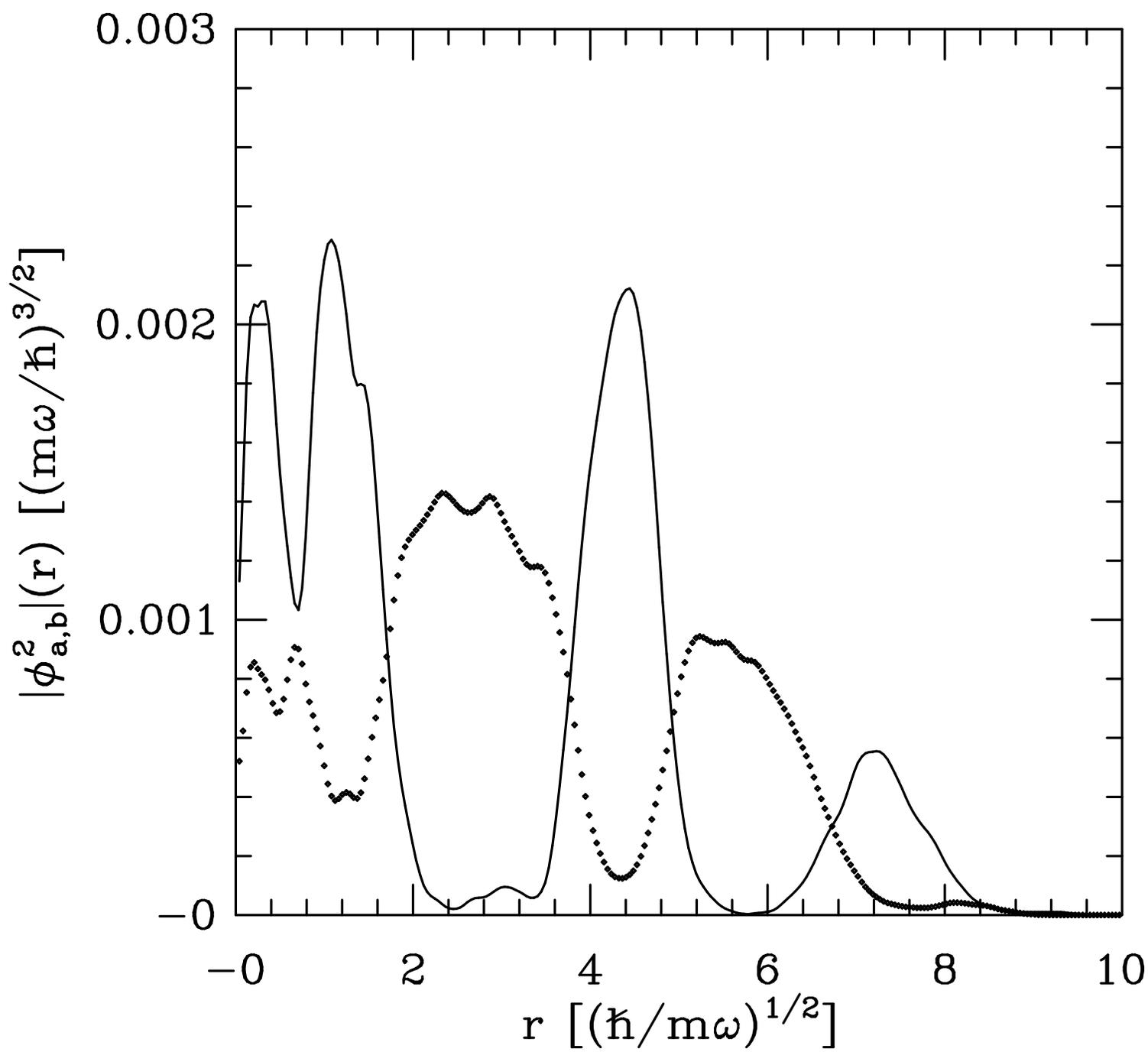
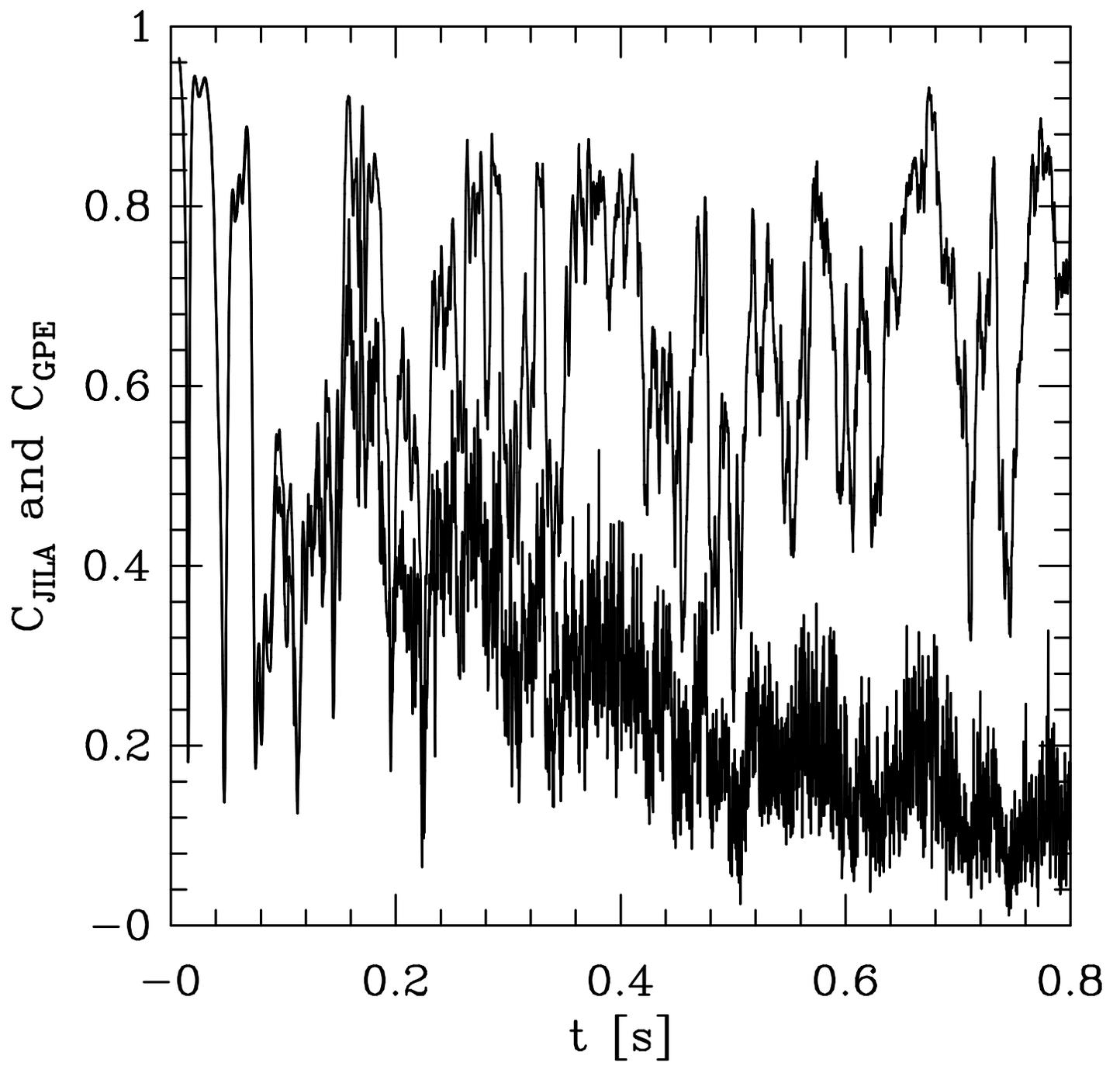
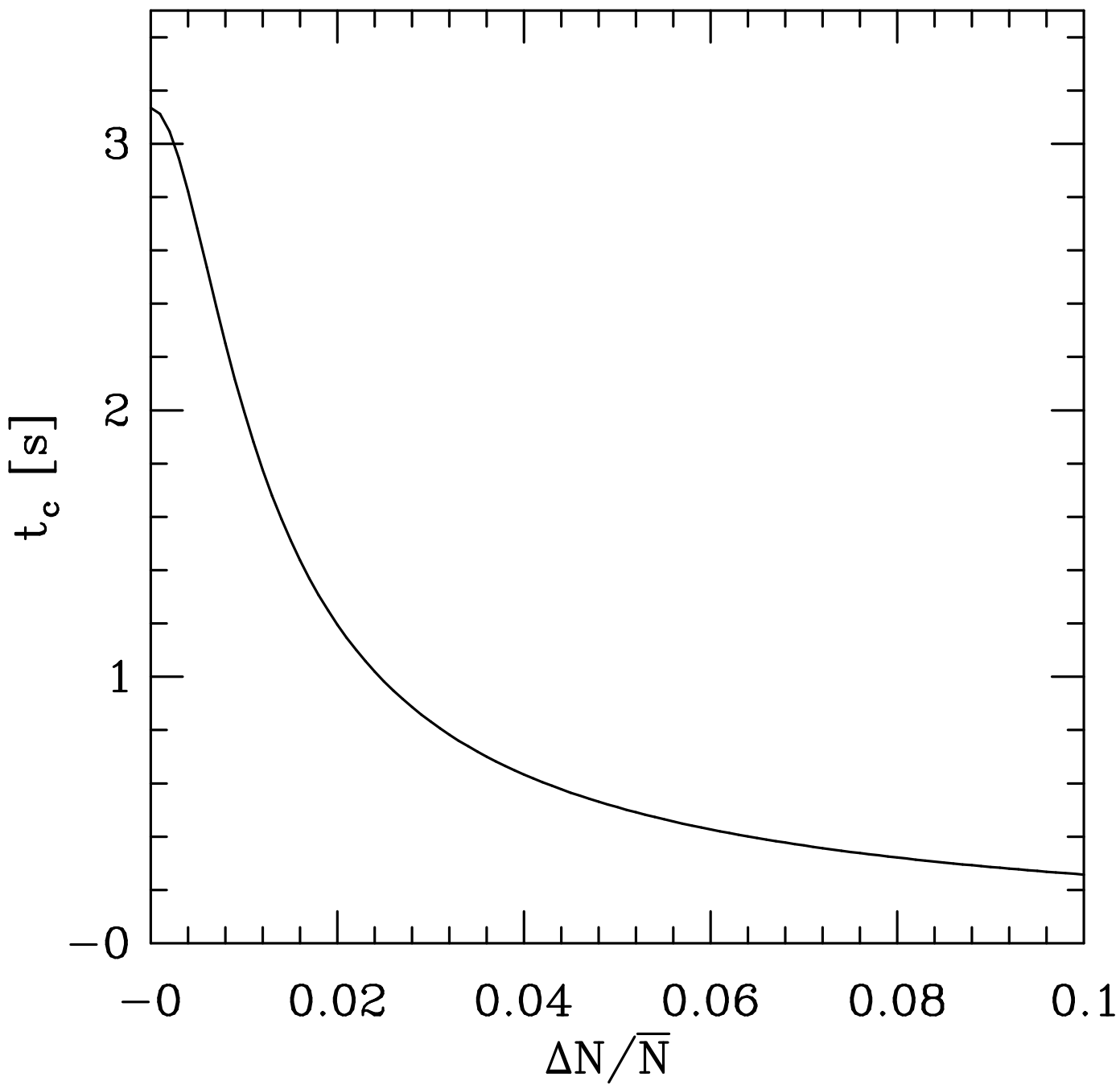
\end{document}